\documentclass[12pt,a4paper]{article}
\usepackage{graphics}
\begin{document}
 
 \thispagestyle{empty}
 
 \title{Invariant operator due to F. Klein
 quantizes\\ H. Poincar\'{e}'s dodecahedral 3-manifold.}
 \author{Peter Kramer,\\
 Institut f\"ur Theoretische Physik der\\
 Universit\"at  72076 T\"ubingen, Germany}
 \maketitle

\section*{Abstract.}
The eigenmodes of the Poincar\'e dodecahedral 3-manifold $M$ are 
constructed  as  eigenstates of a novel invariant operator. 
The 
topology of $M$ is characterized by the 
homotopy group $\pi_1(M)$, given by loop composition on $M$, 
and by the isomorphic
group of deck transformations $deck(\tilde{M})$, acting 
on the universal cover $\tilde{M}$. ($\pi_1(M)$, $\tilde{M}$) are known 
to be the binary icosahedral group ${\cal H}_3$ and the sphere $S^3$
respectively.  
Taking $S^3$  as the group manifold $SU(2,C)$ it is shown that 
$deck(\tilde{M}) \sim {\cal H}^r_3$ acts on $SU(2,C)$ 
by right multiplication.
A semidirect product group is constructed 
from ${\cal H}^r_3$ as normal subgroup and from a second
group ${\cal H}^c_3$ which provides the icosahedral symmetries of $M$.
Based on F. Klein's fundamental 
icosahedral ${\cal H}_3$-invariant,  
we construct a novel hermitian ${\cal H}_3$-invariant 
polynomial (generalized Casimir) operator ${\cal K}$. Its
eigenstates with  eigenvalues $\kappa$ quantize  
a complete orthogonal basis 
on Poincar\'{e}'s dodecahedral 3-manifold. The eigenstates of lowest
degree $\lambda=12$ are 12 partners of Klein's invariant polynomial.
The analysis has applications in cosmic topology \cite{LA},\cite{LE}.
If the Poincar\'{e} 3-manifold $M$ is assumed  
to model the space part of a cosmos, the observed temperature
fluctuations of the cosmic microwave background
must admit an expansion in eigenstates of ${\cal K}$.

.

\section{Introduction.}

We  present a rigorous Lie algebraic operator  
appoach to the eigenmodes of the Poincar\'{e}
dodecahedral 3-manifold.
The eigenmodes are replaced by eigenstates of an  
operator ${\cal K}$ invariant under deck transformations. They are  
quantized by their eigenvalues. These eigenstates 
on $M$ form a complete orthogonal basis appropriate for the expansion of 
observables.
 
(I): Given a topological manifold $X$, the fundamental or first homotopy group
$\pi_1(X)$ is formed by loops and their composition. There is a manifold
$\tilde{X}$, called the universal cover of $X$, which is simply connected
and on which a group $deck(\tilde{X})$ isomorphic to $\pi_1(X)$
acts discontinuously by so-called deck transformations. The quotient set 
$\tilde{X}/deck(\tilde{X})$ is the original $X$. For $X=M$ being the Poincar\'{e}
dodecahedral 3-manifold, the universal cover is the 3-sphere 
$\tilde{M}=S^3$, and the 
fundamental group is $\pi_1(M)={\cal H}_3$, the binary icosahedral group. 
We use the notation ${\cal H}_3$ to avoid 
confusion with the symbol $H_3$ \cite{HU} p. 46 for the icosahedral Coxeter group. 

(II): The functional analysis on these topological manifolds starts on the universal 
cover and employs representation theory.  As eigenmodes of $S^3$ we take 
the spherical harmonics,
homogeneous solutions of fixed degree $\lambda$ of the Laplace equation .
They transform according to  particular irreducible representations 
(irreps)
$D^j \times D^j,\; j=\lambda/2$ of the full group of isometries $SO(4,R)$
of $S^3$. 
The deck transformations  form a subgroup ${\cal H}_3 <SO(4,R)$
of isometries. We employ the
subduction of 
irreps under the restriction $SO(4,R)>{\cal H}_3$
to decompose the spherical harmonics on $S^3$ into modes 
transforming under the irreps $D^{\alpha}$ of ${\cal H}_3$.
This subduction is solved with a Lie-algebraic novel operator technique.
Among these modes we determine the subset transforming according to the 
identity
irrep $D^{\alpha_0}$ of ${\cal H}_3$. Taken as functions on $S^3$ they are 
${\cal H}_3$-periodic with respect to the decomposition of
$S^3$ into  copies of
$M$. Their unique restrictions from $S^3$ to $M$ are the eigenmodes
of $M$. 

The analysis is carried out as follows: 
In section 2 we set up the continuous geometry of the universal cover $S^3$.
We take $S^3$ in the equivalent form of the group manifold $SU(2,C)$
and implement the action of $SO(4,R)$. 
In section 3 we 
introduce discrete  subgroups of $SO(4,R)$ which  provide 
all the discrete group actions needed.  
In section 4 we show that    
the  group of deck transformations $deck(\tilde{M}) \sim {\cal H}_3$
is the normal subgroup in a semidirect product, Lemma 2.
The elements of  ${\cal H}_3$ correspond to Hamilton's  icosians.
We show that $deck(\tilde{M}) = {\cal H}_3$ 
acts on $SU(2,C)$ by right action, Lemma 3, and interprete
the semidirect product in terms of symmetries of $M$, Lemma 4.
The spherical harmonics 
on $S^3$ are identified in section 5 as Wigner's $D$-functions 
on $SU(2,C)$, Lemma 5. 
>From Felix Klein's fundamental icosahedral invariant \cite{KL} in section 6 we
obtain   12 more ${\cal H}_3$-invariant polynomials of degree $12$. 
They are classified in terms of left action generators in Lemma 7.
One of these invariant polynomials due to Klein allows to pass  
in sections 7, 8 to a invariant hermitian generalized 
Casimir operator ${\cal K}$ on the enveloping Lie algebra of 
$SO(4,R)$, Lemma 8. It characterizes 
the irrep subduction $SO(4,R)>{\cal H}_3$. The operator is shown 
to quantize the 
${\cal H}_3$-invariant eigenmodes of the Poincar\'{e} 3-manifold $M$ 
as eigenstates, 
Theorem 1. The term quantization is used here in the  sense given in  
Schr\"{o}dinger's   papers \cite{SC} 
on quantization as an eigenvalue problem.

We analyze the spectrum of ${\cal K}$
and completely resolve its degeneracy by additional
hermitian operators and quantum numbers. The eigenstates of ${\cal K}$ 
up to $j=6$ are
explicitly given in algebraic form.
Analysis of the spectrum of ${\cal K}$ 
yields the selection rules for eigenmodes on $M$ versus 
those on $\tilde{M}=S^3$. 
A comparison to related work from cosmic topology 
is given in section 11.
Operator symmetrizations appearing in ${\cal K}$ are
carried out in the Appendix.

\section{Continuous geometry of $S^3$.}

The sphere $S^3$, embedded in Euclidean space $E^4$ as a manifold, 
is in one-to-one correspondence to the group manifold $SU(2,C)$.
Let $(x_0,x_1,x_2,x_3)\in E^4$ be orthogonal coordinates and define
\begin{eqnarray}
\label{e1}
u(x)&:=& 
\left[
\begin{array}{ll}
x_0-ix_3& (-i)(x_1-ix_2)\\
(-i)(x_1+ix_2)&(x_0+ix_3)
\end{array}
\right]
=x_0\sigma_0 +(-i)\sum_{j=1}^3 x_j\sigma_j,
\\ \nonumber
&&{\rm det}(u(x))= \sum_0^3 (x_i)^2=1.
\end{eqnarray}
We shall employ complex variables and rewrite eq. \ref{e1} as
\begin{eqnarray}
\label{e2}
u(x)=u(z,\overline{z})=
\left[
\begin{array}{ll}
z_1 &z_2 \\
-\overline{z}_2&\overline{z}_1,
\end{array}
\right],\; z_1\overline{z}_1+z_2\overline{z}_2=1.
\end{eqnarray}
In eq. \ref{e1}, $\sigma_0$ denotes the $2 \times 2$ unit matrix and $\sigma_j$
the Pauli matrices. The four cartesian coordinates in eq. \ref{e1} could 
be replaced by three independent real Euler (half)-angles 
$(\alpha, \beta, \gamma), 0\leq (\alpha,\gamma) <4\pi, 0\leq \beta <2\pi$,
see \cite{ED} pp. 53-67. To fully cover $S^3$, the Euler angular parameters 
must take their full range of values.
Eq. \ref{e1} may be considered
as the physicists version of an expression employing quaternions, 
given in a different context in \cite{MO}, \cite{PA} as 
\begin{eqnarray}
\label{e3}
&&{\bf x}={\bf 1}x_1'+{\bf i}x_2'+{\bf j}x_3'+{\bf k}x_4',
\\ \nonumber
&&{\bf 1}=\sigma_0,\; {\bf i}=i\sigma_3,\; {\bf j}= i\sigma_2,\;
{\bf k}=i\sigma_1,
\\ \nonumber
&& (x_1',x_2',x_3',x_4') \rightarrow (x_0, -x_3, -x_2, -x_1).  
\end{eqnarray}
The second and third line convert the scheme of eq. \ref{e3}, first line,  
to the one used in
eq. \ref{e1}. Any matrix $u(x)  \in SU(2,C)$   is mapped  
one-to-one to  a point on $S^3$  by  eqs. \ref{e1}, \ref{e2}. 

For 
two matrices $u(y),u(x)$ of the
type eq.\ref{e1} we define the hermitian scalar product by
\begin{equation}
\label{e4}
\langle u(y), u(x)\rangle :=\frac{1}{2} {\rm Trace}(u(y)^{\dagger} \times u(x))
= y_0x_0 + \sum_{j=1}^3 y_jx_j.
\end{equation}

Consider group actions of the type 
$SU(2,C) \times SU(2,C) \rightarrow SU(2,C)$.
The left, right, and conjugation actions of $g_l, g_r, g_c  \in SU^{l,r,c}(2,C)$ on  
$u \in SU(2,C)$ are  defined by 
\begin{equation}
\label{e5}
(g_l, u) \rightarrow g_l^{-1}u,\; (g_r, u) \rightarrow ug_r,\; 
(g_c,u) \rightarrow g_c^{-1}ug_c. 
\end{equation}
The left and right action by $SU^{l,r}(2,C)$ commute. When combined they 
yield a 
direct product action $SU^l(2,C) \times SU^r(2,C)$ linear in the 
matrix elements of $u$ 
which is easily shown to fully cover $SO(4,R)$.
Removing the elements $(I,-I):=(\sigma_0, -\sigma_0)$ of the stability group 
$Z_2$ from the action of $SU^l(2,C) \times SU^r(2,C)$ on $SU(2,C)$ 
we have the well-known result 
\begin{equation}
\label{e6}
SO(4,R) = (SU^l(2,C) \times SU^r(2,C))/Z_2.
\end{equation}
 Turn to the conjugation action. By writing $u$ according to eq.
\ref{e1}, one sees that $x_0$ is unchanged while the conjugation of the
Pauli matrices yields
\begin{equation}
\label{e7}
g_c^{-1}\sigma_jg_c= \sum_{i=1}^3 \sigma_iD_{ij}^1(g_c),
\end{equation}
so that $(x_0,x_1,x_2,x_3) \in E^4$ transform according to 
$(1 \times D^1)$,
with $D^1$ the defining real irrep
of the rotation group $SO(3,R)$. 

Combining right multiplication and conjugation,  define a 
multiplication rule
\begin{equation}
\label{e9}
(g_{r_1},g_{c_1}) \times (g_{r_2},g_{c_2})
:=(g_{c_2}^{-1}g_{r_1}g_{c_2}g_{r_2},g_{c_1}g_{c_2}).
\end{equation}
This multiplication rule is associative and 
generates a group. 
Subgroups $SU^r(2,C)$, $SU^c(2,C)$ are generated by the elements
$(e,g_c),\;(g_r,e)$ respectively. One finds the conjugation rule
\begin{equation}
\label{e9a}
(e,g_c)(g_r,e)(e,g_c^{-1})=(g_cg_rg_c^{-1},e),
\end{equation} 
which shows that eq. \ref{e9} yields a semidirect product 
\begin{equation}
\label{e9b}
SU^r(2,C) \times_s  SU^c(2,C),  
\end{equation}
with $SU^r(2,C)$ the normal subgroup. We let this semidirect 
product act on $u \in SU(2,C)$ as
\begin{equation}
\label{e9c}
((g_r,g_c), u) \rightarrow g_c^{-1}ug_cg_r.
\end{equation}
This action, with emphasis on the right action, was the reason for the
choice eq. \ref{e9} and will be used in what follows. 
The action is linear in the matrix elements of $u$ and from
eq. \ref{e2} yields a homomorphism from the 
semidirect product group eq. \ref{e9b}
to $SO(4,R)$. Since $(g_r,\pm g_c)$ yield the same element of $SO(4,R)$,
the homomorphism is two-to-one, and from the stability group under
the action eq. \ref{e9c} we find 
\begin{equation}
\label{e10}
SO(4,R) = (SU^r(2,C) \times_s SU^c(2,C))/Z_2.
\end{equation}

>From the group actions we infer the relevant representation theory:
The correspondence of $SO(4,R)$ to the direct product eq. \ref{e6} 
implies that its general irreducible representations 
(irreps)
are the direct products $D^{j_1} \times D^{j_2}$ of two independent irreps
of $SU^{l,c}(2,C)$. The sphere $S^3$ can be obtained by acting with $SO(4,R)$ on the 
representative point $(1,0,0,0)$. This point has the stability group
$SO(3,R)$, acting according to eq. \ref{e7}. The stability group
corresponds to the
second factor (conjugation action) of the semidirect product action 
eq. \ref{e9c}. It 
then follows that the 
homogeneous or coset space has the 
structure 

\begin{equation}
\label{e11}
S^3:= SO(4,R)/SO(3,R)=SU(2,C),
\end{equation}
in line with eq. \ref{e1} and 
corresponding to the 
first factor (right action) in the semidirect product eq. \ref{e9b}.

In relation with the Laplace equation on $E^4$ we are interested in 
the spherical harmonics on $S^3$ and their transformation under $SO(4,R)$.
They will be identified as the Wigner representation functions
$D^j(u),\; u \in SU(2,C)$ in section 5 and shown to transform 
according to the irreps   
$D^{j} \times D^{j}$ of $SU^l(2,C) \times SU^r(2,C)$.

\section{Discrete and  Coxeter groups acting on $S^3$.}

Coxeter groups \cite{HU} chapter 5 are finitely generated by 
reflections such that any product
of two reflection generators is of finite order.
We shall show in the next sections that the discrete
group of deck transformations and the symmetry group 
of the dodecahedral Poincar\'{e}
3-manifold appear as subgroups of a spherical Coxeter group.

Consider the spherical Coxeter group  \cite{HU} with 
four generators $R_1,R_2,R_3,R_4$, Coxeter-Dynkin diagram
and relations
\begin{eqnarray}
\label{e12}
&&\circ \frac{5}{}\circ \frac{3}{}\circ 
\frac{3}{}\circ
\\ \nonumber 
&&:=\langle  R_1,R_2,R_3,R_4|
\\ \nonumber
&&(R_1)^2=(R_2)^2=(R_3)^2=(R_4)^2=
(R_1R_2)^5=(R_2R_3)^3=(R_3R_4)^3=e \rangle.
\end{eqnarray}

This Coxeter group admits an isometric action on $E^4$  
by Weyl reflections 
\begin{equation}
\label{e13}
W(a_i)x := x-2\frac{\langle x,a_i\rangle}{\langle a_i,a_i\rangle}\; a_i   
\end{equation}
in four hyperplanes perpendicular to  four
Weyl vectors $a_1,\ldots, a_4$. The action maps $S^3$ to $S^3$. We choose these vectors as
\begin{eqnarray}
\label{e14}
&&\begin{array}{llllll}
a_1& =& (0,& 0,                         &1,           &0),\\
a_2& =& (0,& -\frac{\sqrt{-\tau+3}}{2},&\frac{\tau}{2},&0),\\
a_3& =& (0,&-\sqrt{\frac{\tau+2}{5}},& 0,&-\sqrt{\frac{-\tau+3}{5}}),\\ 
a_4& =& (\frac{\sqrt{2-\tau}}{2},&0,& 0,&-\frac{\sqrt{\tau+2}}{2}),
\end{array}
\\ \nonumber
&&\langle a_i, a_i \rangle=1, i=0,\ldots ,3,
\; \langle a_1, a_2 \rangle=\cos(\pi/5)=\tau/2,
\\ \nonumber
&&\langle a_2, a_3 \rangle=\cos(\pi/3)=1/2,
\;  \langle a_3, a_4 \rangle=\cos(\pi/3)=1/2,\;
\\ \nonumber 
&&\langle a_i,a_j \rangle =0\; {\rm otherwise},\; \tau=(1+\sqrt{5})/2.
\end{eqnarray}
Any Weyl reflection $W$ in the Coxeter group eq. \ref{e12}
is represented by a $4 \times 4$ real matrix of 
determinant $det(W)=-1$ and so does not belong to $SO(4,R)$.
We shall work with the normal subgroup of the Coxeter group given by 
\begin{equation}
\label{e14a}
S(\circ \frac{5}{}\circ \frac{3}{}\circ 
\frac{3}{}\circ):= (\circ \frac{5}{}\circ \frac{3}{}\circ 
\frac{3}{}\circ)\cap SO(4,R).
\end{equation}
All elements of this normal
subgroup contain an even number of reflections when written as 
products of generators. They can be generated from the  products
$(R_iR_{i+1}), i=1,2,3$. 

The subgroup 
\begin{equation}
\label{e15}
\circ \frac{5}{}\circ \frac{3}{}\circ 
=\langle R_1,R_2,R_3| (R_1)^2=(R_2)^2=(R_3)^2=(R_1R_2)^5=(R_2R_3)^3=e\rangle
\end{equation}
is the icosahedral
Coxeter subgroup $H^3$ including reflections and of order $120$. 
Any element $g$ of this group  leaves $x_0$ unchanged and acts 
on $E^4$ as 
$(1\times D^1(g))$.

The even normal subgroup of the icosahedral Coxeter group eq. \ref{e15}
we denote by 
\begin{equation}
\label{e15a}
S(\circ \frac{5}{}\circ \frac{3}{}\circ)
:= (\circ \frac{5}{}\circ \frac{3}{}\circ)\cap SO(4,R),
\end{equation}
it is of order
$60$ and consists of all icosahedral rotations.
In particular  
the even element
$(R_1R_2)$ stabilizes the vector $(0,0,0,1) \in E^4$.
Given any element $g_c$ of the binary icosahedral group ${\cal H}^c_3$,
its conjugation map eq. \ref{e5} on $E^4$ yields a two-to-one homomorphism
$(1,g_c) \rightarrow (1 \times D^1(g_c)) \in 
S(\circ \frac{5}{}\circ \frac{3}{}\circ)
< SO(4,R)$ which fully covers
the icosahedral group eq. \ref{e15a}. 

Restricting the two groups in the continuous semidirect product 
eq. \ref{e9b}
with multiplication rule eq. \ref{e9} to two binary  icosahedral 
groups, we define a semidirect product 
\begin{equation}
\label{e15b}
{\cal H}^r_3 \times_s {\cal H}^c_3,
\end{equation}
with normal subgroup
$ {\cal H}^r_3$ and homomorphic to an $SO(4,R)$ 
action eq. \ref{e9c} on $S^3$. 

{\bf Lemma 1: Discrete groups acting on $S^3$}:
The semidirect product group eq. \ref{e15b} admits a 
two-to-one homomorphism to $S(\circ \frac{5}{}\circ \frac{3}{}\circ 
\frac{3}{}\circ)$ eq. \ref{e14a}. 

{\em Proof}: The group $S(\circ \frac{5}{}\circ \frac{3}{}\circ 
\frac{3}{}\circ)$ may be generated by the even products 
$(R_1R_2, R_2R_3, R_3R_4)$. Preimages in 
${\cal H}^r_3 \times_s {\cal H}^c_3$ of the first two products
can be constructed from elements $(e,g_c) \in {\cal H}^c_3$. 
For $(R_3R_4)$ it suffices to construct 
the preimage of some even element
$(QR_4), Q \in \circ \frac{5}{}\circ \frac{3}{}\circ$,
since then $(R_3R_4)= (R_3Q^{-1})(QR_4),\; 
(R_3Q^{-1}) \in S(\circ \frac{5}{}\circ \frac{3}{}\circ)$.
Such a preimage from ${\cal H}^r$ will be constructed 
in Lemma 2 below.

\section{The homotopy group and the group of deck transformations 
of the Poincar\'{e} 3-manifold $\tilde{M}$.}

For general topological notions we refer to 
\cite{SE1}, \cite{MU} and \cite{TH}.
The topology of a manifold is 
characterized by its
homotopy and homology groups. It is well known, \cite{SE1} p. 217, that the 
first homology
groups  of $M$ and  $S^3$ are trivial and therefore
fail to discriminate the topologies of these two manifolds. The homotopy group
$\pi_1(M)$ acts by loop composition. It was constructed and analyzed
by Seifert and Threlfall \cite{SE2} pp. 216-218, 
using loops  
along the edges of the dodecahedron $M$,
and shown by Threlfall in \cite{TH2} from its generators and 
relations 
to be isomorphic to the 
binary icosahedral group ${\cal H}_3$ of order $|{\cal H}_3|=120$. 
The order comes about
by the two-to-one mapping between the binary and the proper 3-dimensional 
icosahedral group of order $60$.
The group ${\cal H}_3$ is interpreted 
geometrically in \cite{TH2}.

To begin with we construct on $S^3$ a spherical dodecahedron $M$.
Consider the Weyl vector $a_4$ from eq. \ref{e14}. The 
Weyl hyperplane  
$\langle x, a_4 \rangle=0$ in $E^4$, invariant under the Weyl reflection 
$W(a_4)$, intersects $S^3$ in a unit sphere $S^2$. The vector
$a_4$ and its Weyl hyperplane under the icosahedral rotation group 
$S(\circ \frac{5}{}\circ \frac{3}{}\circ)$ each have $12$ images
in $E^4$ and $12$ corresponding unit spheres $S^2$ on $S^3$. These twelve
unit spheres bound a convex spherical dodecahedron $M$ on $S^3$.
We label the 12 faces of the spherical dodecahedron $M$ as follows: 
The spherical face in the hyperplane perpendicular to $a_4$ 
we denote as $\partial_1M$, the five faces 
sharing an edge with $\partial_1M$ we label counterclockwise by $\partial_2M,
\ldots, \partial_6M$, and 
the faces opposite to $\partial_1M,\ldots , \partial_6M$ by 
$\partial_{\overline{1}}M, \ldots, \partial_{\overline{6}}M$.
The spherical dodecahedron differs from the Euclidean 
dodecahedron, compare \cite{TH} p.35: 
The faces are
spherical, faces adjacent to an edge have dihedral angle $2\pi/3$, any
edge becomes a 3fold rotation axis generated by $(R_3R_4)$ or its 
conjugates.
In the next equation we relate the enumeration of faces \cite{KR1}, \cite{KR2} to
the letters used in \cite{SE1}, \cite{SE2}:
\begin{equation}
\label{e16}
\begin{array}{lllllll}
\cite{SE1},\cite{SE2}:&A&B&C&D&E&F.\\
\cite{KR1}:           &1&5&6&2&3&4.\\
\end{array}
\end{equation}
Since the icosahedral Coxeter group eq. \ref{e15} 
permutes the faces of $M$,
the generators of $\circ \frac{5}{}\circ \frac{3}{}\circ$ 
can be written as signed permutations in cycle 
notation as follows \cite{KR1}, \cite{KR2}:
\begin{equation}
\label{e17}
R_1=(23)(46), R_2=(24)(56), R_3=(15)(2\overline{3}).
\end{equation}
We multiply permutations from right to left. 
The icosahedral rotations of the group 
$S(\circ \frac{5}{}\circ \frac{3}{}\circ)$ eq. \ref{e15a}
provide the symmetry of the dodecahedron $M$.

The classical prescription for constructing from the spherical
dodecahedron the Poincar\'{e} dodecahedral 3-manifold $M$, given 
by Weber, Seifert and Threlfall in \cite{SE1},\cite{SE2}, 
is the gluing of opposite faces 
of the dodecahedron after a rotation by an angle $\pi/5$.

\begin{center}
\includegraphics{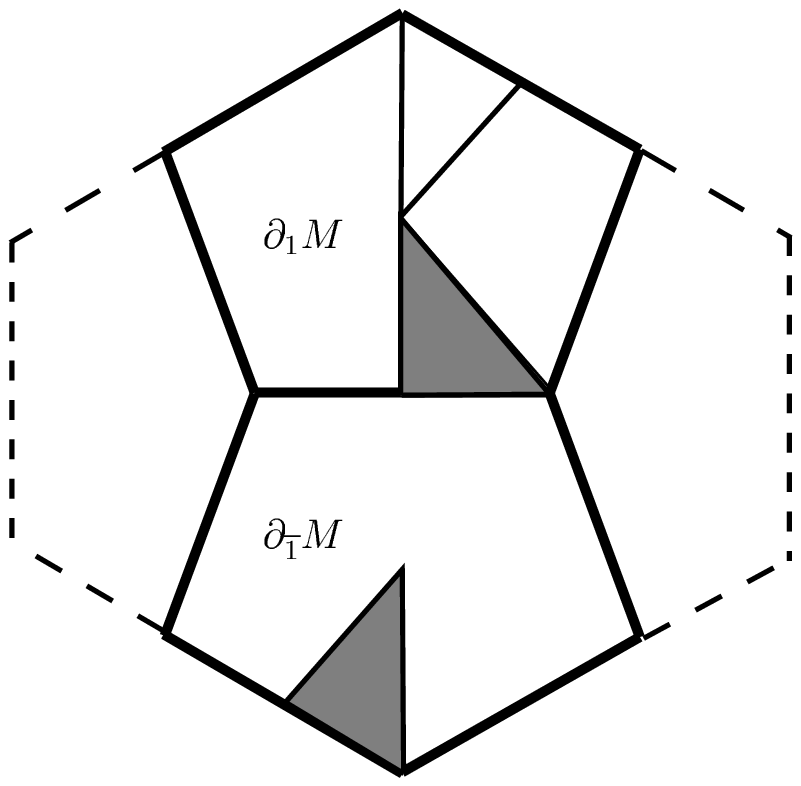}
\end{center}

Fig. 1. Face-to-face gluing for the dodecahedral Poincar\'{e} 
manifold $M$ as a product  of three operations from the Coxeter group
eq. \ref{e12}:  Two opposite pentagonal faces of the dodecahedron $M$
(dashed lines) are shown in a projection along a 2-fold axis perpendicular to the figure. 
(1) The shaded triangle of face
$\partial_{\overline{1}}M$, bottom pentagon, is mapped by the inversion $P$ to the white
triangle on the opposite face $\partial_1M$, top pentagon. 
(2) The white triangle is mapped by a counterclockwise 
rotation $5_1^{-2}$ by $-4\pi/5$ about the axis passing through 
the midpoints of both faces  
to its shaded final position. (3) The Weyl generator  
$R_4$ reflects $M$ in the 
hypersurface containing face $\partial_1M$ and maps $M$  to its face-to-face 
neighbour $R_4M$. The product $C_1 =R_4\, 5_1^{-2}\,P$ reproduces 
the counterclockwise rotation by $\pi/5$, shift between opposite faces, and 
gluing prescribed in \cite{SE1}, \cite{SE2} as a  generator of the 
homotopy group.
 
\vspace{0.2cm}

The group of deck transformations $deck(\tilde{X})$ \cite{SE1} pp. 196-197,
\cite{MU} p. 398
is defined as the group $G$ acting on the universal cover $\tilde{X}$ 
such that $\tilde{X}/deck(\tilde{X})$ equals $X$. 
By general theorems
given in \cite{SE1} pp. 181-198, the groups
$deck(\tilde{X})$ and $\pi_1(X)$ 
are isomorphic. In the present case the action of 
$deck(\tilde{M}) \sim {\cal H}_3$  tiles the universal cover 
$\tilde{M}=S_3$
by dodecahedral copies of $M$, produces the face-to-face gluing conditions,
and forms the $120$-cell, \cite{SO} pp. 172-176.

We construct $deck(\tilde{M})$ within the 
geometry of $S^3$ given in section 2 by use of the group 
eqs. \ref{e14a}, \ref{e15b}. By shifting, rotating and gluing a copy  $C_1M$ to the 
face $\partial_1M$ of $M$ we find 
within the groups eqs. \ref{e14a}, \ref{e15b}  
a first generator $C_1 \in deck(\tilde{M})$.  
Denote  the 5-fold rotation around the midpoint of face $\partial_1M$ by 
$5_1=(R_2R_1)=(23456)$ in cycle notation eq. \ref{e17}. 
Consider the following product of
elements from the icosahedral  Coxeter group: the inversion 
$P:=(1\overline{1})(2\overline{2})(3\overline{3}) 
(4\overline{4})(5\overline{5})(6\overline{6})$, followed by a rotation 
$(5_1)^{-2}:=(53642)$,
followed  by $R_4$. 
These operations are illustrated in Fig. 1. Note that 
$P$ commutes with all elements of the icosahedral Coxeter group. We claim:

{\bf Lemma 2: The group of deck transformations  is the
normal subgroup ${\cal H}^r_3$ of the semidirect product 
group eq. \ref{e15b} }: 
The element 
\begin{equation}
\label{e18}
C_1:=R_4\;5_1^{-2}\;P
\end{equation} 
in the 
group eq. \ref{e14a} is the generator $C_1$ of $deck(\tilde{M})$ 
correponding to shifting, rotating $M$ by $\pi/5$, and gluing
face $\partial_{\overline{1}}(C_1M)$ to face $\partial_1M$ in $S^3$.\\ 
{\em Proof}: 
The element $C_1$ of eq. \ref{e18} contains two
reflections $R_4$ and $P$ of determinant $det=-1$ and so belongs
to the normal subgroup eq. \ref{e14a} of the full Coxeter group.
In geometric terms it is clear that the operation 
$C_1$ of eq. \ref{e18} reproduces, in terms of elements of the  
group eq. \ref{e14a}, the prescription for rotating and gluing 
given in \cite{SE1}, \cite{SE2}, see Fig. 1. 

We now analyze  the correspondence of $C_1$ 
eq. \ref{e18} to an element of the group of deck transformations
$deck(\tilde{M}) \sim {\cal H}_3$.
For this purpose we determine the action of 
$g=(R_4, 5_1^{-2},P, C_1)$ 
in the geometry of section 2. 
In terms of mappings of the complex parameters eq. \ref{e2}
and their composition one finds
for the elements $(g,(z_1,z_2))\rightarrow im_g(z_1,z_2)=(z_1',z_2')$
\begin{eqnarray}
\label{e19}
&&\begin{array}{ll}
g:     & (z_1',z_2')=im_g(z_1,z_2),\\
R_4:   & (-\epsilon^2\overline{z}_1, z_2),\\
5_1^{-2}:& (z_1,\epsilon^2z_2),\\
P:       & (\overline{z_1},-z_2),\\
R_4\;5_1^{-2}\;P:&(-\epsilon^{-2}z_1, -\epsilon^2z_2),\\
\end{array},
\\ \nonumber
&& \epsilon:=\exp(2\pi i/5).
\end{eqnarray}
The action of
$C_1$ eq. \ref{e18} on $(z_1,z_2)$ from eq. \ref{e19} 
can be rewritten in terms of a right action on $u(x)$,
\begin{eqnarray}
\label{e21}
&&C_1:\;
u(x) \rightarrow u(x) v,\;
\\ \nonumber
&&v=
\left[
\begin{array}{ll}
-\epsilon^2&0\\
0& -\epsilon^{-2}
\end{array}
\right]
\in {\cal H}^r_3.
\end{eqnarray}
That $v \in {\cal H}^r_3$ follows by comparison with 
Klein's list of elements of the binary icosahedral group 
given in eq. \ref{e23} below.

Any other element of $deck(\tilde{M})$, 
corresponding to the gluing of another pair of 
faces, can be obtained within the semidirect product 
group eq. \ref{e15b}
by conjugation according to eq. \ref{e9a} of 
$v \in {\cal H}^r_3$ eq. \ref{e21} 
with an element $g_c \in {\cal H}^c_3$,
homomorphic to an element of  
the icosahedral group $S(\circ \frac{5}{}\circ \frac{3}{}\circ)$.
In particular one finds $(C_i)^{-1}=C_{\overline{i}}$.
It is a nontrivial result of \cite{SE2}, \cite{TH2} that the
special elements corresponding to $6$ gluings generate 
the binary icosahedral group ${\cal H}^r_3$ isomorphic to 
$deck(\tilde{M})$ acting on $S^3$. Therefore we find

{\bf Lemma 3: The group of deck transformations 
acts from the right on $S^3 \sim SU(2,C)$}:
The group of deck transformations $deck(\tilde{M}) \sim {\cal H}_3$ 
acts on $S^3$ 
in the parametrization
eq. \ref{e1} by right multiplication 
$u(x)\rightarrow ug_r,\; g_r\in {\cal H}^r_3$ with the  elements 
$g_r$ given by Klein as in eq. \ref{e23} below.

{\bf Lemma 4: Symmetry and  group of deck transformations for $M$.} The semidirect product group 
${\cal H}^r_3 \times_s {\cal H}^c_3$ eq. \ref{e9b} is associated
with the Poincar\'{e} dodecahedral 3-manifold $M$. Under the two-to-one
homomorphism ${\cal H}^r_3 \times_s {\cal H}^c_3\rightarrow 
S(\circ \frac{5}{}\circ \frac{3}{}\circ 
\frac{3}{}\circ) <SO(4,R)$, 
the groups ${\cal H}^r_3$ and ${\cal H}^c$
provide the groups of deck transformations and of
icosahedral symmetries  
respectively on $\tilde{M}$.

Klein \cite{KL} pp. 41-42 gives the elements of ${\cal H}_3$ corresponding to
the matrices, compare eq. \ref{e35},
\begin{eqnarray}
\label{e23}
{\cal H}_3:&& S^{\mu},\; S^{\mu}U,\; S^{\mu}TS^{\nu},\;S^{\mu}TS^{\nu}U,\;
\mu, \nu=0,1,2,3,4,\; \epsilon := \exp(2\pi i/5),
\\ \nonumber
&& S:=
\left[
\begin{array}{ll}
\pm \epsilon^3& 0\\
0 & \pm \epsilon^2
\end{array}
\right],\;
U:=
\left[
\begin{array}{ll}
0& \pm 1\\
\mp 1& 0 
\end{array}
\right],\;
T:=
\frac{1}{\sqrt{5}}
\left[
\begin{array}{ll}
\mp (\epsilon-\epsilon^4)& \pm (\epsilon^2-\epsilon^3)\\
\pm (\epsilon^2-\epsilon^3)&\pm (\epsilon-\epsilon^4)
\end{array}
\right].
\end{eqnarray}
Hamilton's  icosians \cite{HA}, \cite{HA2}, 
are the elements of ${\cal H}_3$ in the quaternionic 
basis eq. \ref{e4} of \cite{PA} which agrees with \cite{MO}:
\begin{eqnarray}
\label{e24}
{\cal H}_3: && \frac{1}{2}(\pm 1, \pm 1, \pm 1, \pm 1),
\\ \nonumber
&& (\pm 1, 0,0,0)\: {\rm and}\:{\rm all}\:{\rm permutations},
\\ \nonumber
&&\frac{1}{2}(0, \pm 1, \pm (1-\tau), \pm \tau)
\:{\rm and}\:{\rm all}\:{\rm even}\:{\rm permutations}.
\end{eqnarray}
In terms of the icosahedral group elements lifted to  3-space, 
the set eq. \ref{e23} places a 5-fold axis , eq. \ref{e24} a 2-fold axis in 
the direction 3.  To relate the two sets, construct the 
matrix 
\begin{equation}
\label{e25}
w := 
\left[
\begin{array}{ll}
\cos(\beta/2) & \sin(\beta/2)\\ 
 -\sin(\beta/2)& \cos(\beta/2) 
\end{array}
\right], \;
\cos(\beta)= \sqrt{\frac{\tau+2}{5}}, 
\sin(\beta)= \sqrt{\frac{3-\tau}{5}}.
\end{equation}
This matrix has the conjugation property 
\begin{equation}
\label{e26}
w(\sqrt{\frac{3-\tau}{5}}\sigma_1
+\sqrt{\frac{\tau+2}{5}}\sigma_3)w^{-1}
=\sigma_3.
\end{equation}
which shows that conjugation with $w$ rotates a 2-fold  into a neighbouring 
5-fold icosahedral axis. The matrix $w$ eq. \ref{e25} by conjugation 
relates the  sets eqs. \ref{e23} and \ref{e24} to one another.

\section{Spherical harmonics on $S^3$.}
 
The irreps of $SO(4,R)$ may be characterized by eigenvalues of 
Casimir operators. In second order of the Lie group generators
there are two independent 
Casimir operators, namely the two Casimir operators of 
$SU^l(2,C)$ and $SU^r(2,C)$ whose Lie algebras given in  eqs. \ref{e30a},
\ref{e30b} below commute with one another.  
To characterize the special irreps of $SO(4,R)$ carried by the spherical harmonics
it suffices to use a single second order Casimir operator.
With a short-hand notation $\partial_{y_i}:=\frac{\partial}{\partial y_i}$, we can 
express this second-order Casimir operator of $SO(4,R)$ in terms of 
the Laplace operator, the dilatation operator, and $x^2$,
\begin{eqnarray}
\label{e27}
\Lambda^2:&=& \frac{1}{2}\sum_{i,j=0}^3 
(x_i\partial_{x_j}-x_j\partial_{x_i})^2
\\ \nonumber 
&=& x^2\bigtriangledown^2
-(x \cdot \bigtriangledown)((x \cdot \bigtriangledown)+2).
\end{eqnarray}
We note that the three operators appearing on the right-hand side of 
eq. \ref{e27} form the Lie algebra of the 
symplectic group $Sp(2,R)$.

The spherical harmonics are homogeneous polynomial solutions 
of degree $\lambda$ of the
Laplace equation.
Application of eq. \ref{e27} to solutions $P(x)$ of the Laplace equation 
\begin{equation}
\label{e27a}
\bigtriangleup P(x)=0,\; \bigtriangleup=\bigtriangledown^2, 
\end{equation}
fixes for spherical harmonics 
\begin{eqnarray}
\label{e28}
P(x):\; (x \cdot \bigtriangledown)P(x)&=&\lambda P(x),
\\ \nonumber
\Lambda^2 P(x)=-\lambda(\lambda+2)P(x),
\end{eqnarray} 
the eigenvalue of the second order Casimir operator 
eq. \ref{e27} and the corresponding
irreps of $SO(4,R)$.

In the complex coordinates eq. \ref{e2} we have
\begin{equation}
\label{e29}
(x \cdot \bigtriangledown)=
z_1\partial_{z_1}+z_2\partial_{z_2}+
\overline{z}_1\partial_{\overline{z}_1}+\overline{z}_2\partial_{\overline{z}_2}.
\end{equation}
The generators of the groups $SU^r(2,C), SU^l(2,C)$ acting from the
right and left on $u(z)$ are then found by left and right 
action of the Pauli matrices as
\begin{eqnarray}
\label{e30a}
L^r_{+}&=&\left[z_1\partial_{z_2}-\overline{z}_2\partial_{\overline{z}_1}\right],
\\ \nonumber
L^r_{-}&=&\left[z_2\partial_{z_1}-\overline{z}_1\partial_{\overline{z}_2}\right],
\\ \nonumber
L^r_3&=& (1/2)\left[z_1\partial_{z_1}-\overline{z}_1\partial_{\overline{z}_1}
          -z_2\partial_{z_2}+\overline{z}_2\partial_{\overline{z}_2}\right],
\\
\label{e30b}
L^l_{+}&=&\left[-{z}_2\partial_{\overline{z}_1}
+{z}_1\partial_{\overline{z}_2}\right],
\\ \nonumber
L^l_{-}&=&\left[\overline{z}_2\partial_{{z}_1}
-\overline{z}_1\partial_{{z}_2}\right],
\\ \nonumber
L^l_3&=& (1/2)\left[z_1\partial_{z_1}+z_2\partial_{z_2}
          -\overline{z}_1\partial_{\overline{z}_1}
	  -\overline{z}_2\partial_{\overline{z}_2}\right].
\end{eqnarray}
The left and right generators in eqs. \ref{e30a}, \ref{e30b} 
respectively commute but have among themselves 
the standard $SU(2,C)$ commutation relations
\begin{equation}
\label{e31}
\left[L_3,L_{\pm}\right]= \pm L_{\pm},\; 
\left[L_{+},L_{-}\right]= 2L_3.
\end{equation}
The spherical harmonics are identical to the 
Wigner representation functions $D^j_{m',m}$
of $SU(2,C)$ \cite{ED} pp. 53-67, which in turn are 
equivalent to the  Jacobi polynomials.
>From a generating function \cite{ED} eq. (4.14)
they can be written  
in the notation of eq. \ref{e2}  
as homogeneous polynomials in $(z_1,z_2,\overline{z}_1,\overline{z}_2)$
of degree $2j$,
\begin{eqnarray}
\label{e32}
D^j_{m',m}(z_1,z_2,\overline{z}_1,\overline{z}_2)
&=&
\left[
\frac{(j+m')!(j-m')!}{(j+m)!(j-m)!} \right]^{1/2}
\\ \nonumber
&&\sum_{\sigma}
\frac{(j+m)!(j-m)!}{(j+m'-\sigma)!(m-m'+\sigma)!\sigma!(j-m-\sigma)!}
\\ \nonumber
&&(-1)^{m-m'+\sigma}\; z_1^{j+m'-\sigma}\;
\overline{z}_2^{m-m'+\sigma}\;z_2^{\sigma}
\;\overline{z}_1^{j-m-\sigma},
\\ \nonumber
&&j=0,1/2,1,3/2, \ldots.
\end{eqnarray}
The particular spherical harmonics with $m'=j$ are given from
eq. \ref{e32} by
\begin{equation}
\label{e34}
D^j_{j,m}(z)= \left[\frac{(2j)!}{(j+m)!(j-m)!} \right]^{1/2}
(z_1)^{j+m}(z_2)^{j-m},
\end{equation}
they are analytic in $(z_1,z_2)$.

{\bf Lemma 5: Spherical harmonics on $S^3$ are Wigner $D$-functions}: 
The Wigner $D^j$-functions are homogeneous of degree $\lambda=2j$ and solve
the Laplace equation $\Delta D=0$. The eigenvalues of the operators
$(L^l_3, L^r_3)$ from eqs. \ref{e30a}, \ref{e30b} are $(m',m)$. Under
$SO(4,R)$, the spherical harmonics transform according to
the irreps $D^j \times D^j$ of $SU^l(2,C) \times SU^r(2,C)$.

{\em Proof}: (i) The analytic  $D^j$-functions 
eq. \ref{e34} of degree $2j$ are
easily seen to vanish under the application of the Laplacian $\Delta$.  
All other $D$-functions are obtained by the application of Lie generators
from eqs. \ref{e30a}, \ref{e30b}, which commute with $\Delta$, 
and so they also
must fulfill eqs. \ref{e27a}, \ref{e28}. (ii) Under the left/right actions
$u \rightarrow g_l^{-1}ug_r$, the linear decomposition of
$D^j(g_l^{-1}ug_r)$ in terms of $D^j(u)$ yields $D^j(g_l^{-1}) \times
D^j(g_r)$ as coefficients.

\section{Action of ${\cal H}_3$ and polynomial invariants of degree $\lambda=12$.}

The points of $S^3 \sim SU(2,C)$ are specified by the two complex
numbers $(z_1,z_2)$ in the top row of eq. \ref{e2}. 
The general right action of $SU^r(2,C)$ on these two complex variables 
$(z_1,z_2)$ of $S^3 \sim SU(2,C)$ and on
their tensor products read
\begin{eqnarray}
\label{e35}
&&(z_1',z_2')
= (z_1,z_2) 
\left[
\begin{array}{ll}
a &b\\
-\overline{b}&\overline{a}
\end{array}
\right],
\\ \nonumber
&&(\sqrt{2}z_1'\overline{z}_2',z_1'\overline{z}_1'-z_2'\overline{z}_2',
\sqrt{2}\overline{z}_1'z_2')
\\ \nonumber
&& =(\sqrt{2}z_1\overline{z}_2,z_1\overline{z}_1-z_2\overline{z}_2,
\sqrt{2}\overline{z}_1z_2)
\left[
\begin{array}{lll}
aa &-\sqrt{2}ab&-bb\\
-\sqrt{2}a\overline{b}&a\overline{a}-b\overline{b}&\sqrt{2}\overline{a}b\\
-\overline{b}\overline{b}
&-\sqrt{2}\overline{a}\overline{b}&\overline{a}\overline{a}
\end{array}
\right].
\end{eqnarray}
Eq. \ref{e35} shows that the polynomials $(z_1,z_2)$ 
form a basis of the irrep $D^{1/2}$ of $SU^r(2,C)$.

Felix Klein in his monograph \cite{KL} on the
icosahedral group lets the binary group ${\cal H}_3$ 
with elements eq. \ref{e23} act on two complex variables
$(z_1,z_2)$ in line with eq. \ref{e35}. From a linear fractional transform 
of the complex variables he in \cite{KL} pp. 32-34 passes to real 
cartesian coordinates $(\xi,\eta,\zeta)$. In terms of eq. \ref{e35}
Klein's correspondence  may be written as 
\begin{equation}
\label{e36}
(\sqrt{\frac{1}{2}}(\xi+i\eta),\zeta,\sqrt{\frac{1}{2}}(\xi-i\eta))=
(\sqrt{2}z_1\overline{z}_2,z_1\overline{z}_1-z_2\overline{z}_2,
\sqrt{2}\overline{z}_1z_2),\; \xi^2+\eta^2+\zeta^2=1.
\end{equation}
It is then easily  verified from eq. \ref{e35} that the action of $SU(2,C)$
on $(\xi,\eta,\zeta)$ reproduces the standard group homomorphism 
$SU(2,C) \rightarrow SO(3,R)$. 

In the present analysis, from Lemma 3 we infer that
the group of deck transformations ${\cal H}_3$ acts on $S^3 \sim SU(2,C)$
eq. \ref{e2} from the right as 
\begin{equation}
\label{e37}
(z_1,z_2) \rightarrow (z_1,z_2) g_r,\; g_r \in {\cal H}_3 < SU^r(2,C).
\end{equation}

In \cite{KL} p. 56 eq.(55) Klein derives the  
homogeneous polynomial
\begin{equation}
\label{e38}
f_k(z_1,z_2):= (z_1z_2)\left[ (z_1)^{10} +11(z_1)^5(z_2)^5- (z_2)^{10}\right].
\end{equation}
By construction, Klein's  fundamental polynomial $f_k$ eq. \ref{e38} 
is ${\cal H}_3$-invariant or transforms according to the identity irrep 
$D^{{\alpha}_0}$
of ${\cal H}_3$, is analytic in $(z_1,z_2)$, and forms  the 
starting point of what Klein calls the icosahedral equation.
We emphasize that the explicit form and the invariance of eq. \ref{e38} are
valid if the elements of ${\cal H}_3$ are taken as in eq. \ref{e23}.

Comparing the spherical harmonics eq. \ref{e32},
the  polynomial eq. \ref{e38} up to normalization may be written, in anticipation of
eqs. \ref{e41}, \ref{e42} below, as  
$D^6_{6,\alpha_0}(z_1,z_2)$, has  
degree $2j=12$, $m'=6$, and is a superposition of spherical harmonics
eq. \ref{e34} with $m=-5,0,5$.

It follows from Lemma 3  that the left action of $SU^l(2,C)$
on $S^3 \sim SU(2,C)$ and hence its
generators commute with the right action of the group of 
deck transformations ${\cal H}_3$.
Therefore when we apply powers  $(L_-^l)^r,\; r=0,\ldots, 12$ of the left 
lowering operator 
from eq. \ref{e30b} to the invariant eq. \ref{e38},
we obtain altogether $13$ invariant polynomials 
$D^6_{m,\alpha_0},\; m=6,\ldots,-6$. 
The first seven ones are listed in {\bf Table 1}.

\begin{table}
$
\vspace{0.2cm}
\begin{array}{lll}
m'&  D^6_{m',\alpha_0}   \\
  & \\
6 & z_1^{11}z_2+11z_1^6z_2^6-z_1z_2^{11}\\
  & \\
5 & z_1^{10}\left[-z_1\overline{z}_1+11 z_2\overline {z}_2\right]\\
  & +11\cdot 6 \cdot z_1^5z_2^5\left[-z_1\overline{z}_1+z_2\overline{z}_2\right]\\
  & -z_2^{10}\left[-11z_1\overline{z}_1+z_2\overline {z}_2\right]\\
  & \\
4 &-22 \cdot z_1^9\overline{z}_2
    \left[z_1\overline{z}_1-5z_2\overline{z}_2\right]\\
  & +11 \cdot 6 \cdot z_1^4z_2^4\left[5((z_1\overline{z}_1)^2+(z_2\overline{z}_2)^2)
    -12z_1\overline{z}_1z_2\overline{z}_2 \right]\\
  & -22 \cdot \overline{z}_1z_2^9
    \left[5z_1\overline{z}_1-z_2\overline{z}_2\right]\\
  & \\
3 &-11\cdot 30 \cdot z_1^8\overline{z}_2^2
   \left[z_1\overline{z}_1-3z_2\overline{z}_2\right]\\
  &-11\cdot 60 \cdot z_1^3z_2^3\left[2((z_1\overline{z}_1)^3-(z_2\overline{z}_2)^3)
  -9((z_1\overline{z}_1)^2z_2\overline{z}_2
  -(z_1\overline{z}_1(z_2\overline{z}_2)^2)\right]\\
  & +11 \cdot 30 \cdot \overline{z}_1^2z_2^8
    \left[3z_1\overline{z}_1-z_2\overline{z}_2\right]\\
  & \\
2 &-11\cdot 360 \cdot z_1^7\overline{z}_2^3
    \left[z_1\overline{z}_1-2z_2\overline{z}_2\right]\\
  &+11\cdot 360 \cdot z_1^2z_2^2\left[(z_1\overline{z}_1)^4+(z_2\overline{z}_2)^4
   -8((z_1\overline{z}_1)^3z_2\overline{z}_2
   +z_1\overline{z}_1(z_2\overline{z}_2)^3)
  +15(z_1\overline{z}_1)^2(z_2\overline{z}_2)^2
  \right]\\
  &-11\cdot 360 \cdot \overline{z}_1^3z_2^7
   \left[2z_1\overline{z}_1-z_2\overline{z}_2\right]\\
  & \\
1 &-11\cdot 720  \cdot z_1^6\overline{z}_2^4
   \left[5z_1\overline{z}_1-7z_2\overline{z}_2\right]\\
  &-11\cdot 720 \cdot z_1z_2\\
  &\left[(z_1\overline{z}_1)^5-(z_2\overline{z}_2)^5
  -15((z_1\overline{z}_1)^4z_2\overline{z}_2
  -z_1\overline{z}_1(z_2\overline{z}_2)^4)
  +50((z_1\overline{z}_1)^3(z_2\overline{z}_2)^2
  -(z_1\overline{z}_1)^2(z_2\overline{z}_2)^3)\right]\\
  &+11\cdot 720 \cdot \overline{z}_1^4z_2^6
    \left[7z_1\overline{z}_1-z_2\overline{z}_2\right]\\
  & \\
0 & -11\cdot 42 \cdot 720\;  \cdot z_1^5\overline{z}_2^5 
     \left[z_1\overline{z}_1-z_2\overline{z}_2\right]\\
  &+11\cdot 720\cdot ((z_1\overline{z}_1)^6+(z_2\overline{z}_2)^6
  -36((z_1\overline{z}_1)^5z_2\overline{z}_2
   +z_1\overline{z}_1(z_2\overline{z}_2)^5)\\
  &+225\cdot((z_1\overline{z}_1)^4(z_2\overline{z}_2)^2
  +(z_1\overline{z}_1)^2(z_2\overline{z}_2)^4)
  -400\cdot(z_1\overline{z}_1)^3(z_2\overline{z}_2)^3)\\
  &-11\cdot 42\cdot 720  \cdot \overline{z}_1^5z_2^5
  \left[z_1\overline{z}_1-z_2\overline{z}_2\right]\\        
\end{array}
\vspace{0.2cm}
$
\caption{ ${\cal H}_3$-invariant polynomials $D^j_{m',\alpha_0}$ 
(unnormalized), obtained from Klein's analytic invariant
eq. \ref{e38} by application of powers 
$(L_-^l)^r,\; r=0,\ldots ,6;\; m'=6-r$ of the left 
lowering operator $L^l_-$ from eq. \ref{e30b}.}

\end{table}

The polynomials in {\bf Table 1} may be normalized by introducing the orthonormal basis
eq. \ref{e32}.

Among these invariant polynomials we look for one that, by the Klein  
correspondence eq. \ref{e36}, may be expressed in terms of 
$(\xi, \eta, \zeta)$. This will allow us in the section 8 to pass 
to an invariant in the enveloping Lie algebra.  
A necessary and sufficient condition for rewriting a polynomial 
$P(z_1,z_2,\tilde{z}_1,\tilde{z}_2)$ in terms of   the Klein 
correspondence is that the complex variables
$(z_1,z_2)$ and their complex conjugates must appear with equal powers.
Expressed in terms of the generator  $L_3^l$ of eq. \ref{e30b}, the 
condition requires the eigenvalue $m'=0$ of this generator.
This condition singles out  
from {\bf Table 1} the invariant polynomial 
$D^6_{0,\alpha_0}$. Upon introducing  the three 
powers $(z_1\overline{z}_1-z_2\overline{z}_2)^p,\; p=2,4,6$, this 
invariant polynomial  can be   rewritten as
\begin{eqnarray}
\label{e39}
&&(L^l_-)^6f_k := 11\cdot 6!\;  {\cal K}',
\\ \nonumber
&&{\cal K}'= 
-42((z_1\overline{z}_2)^5+(\overline{z}_1z_2)^5)
(z_1\overline{z}_1-z_2\overline{z}_2) 
\\ \nonumber
&&+(z_1\overline{z}_1-z_2\overline{z}_2)^6
\\ \nonumber
&&-30 (z_1\overline{z}_2)(\overline{z}_1z_2)
(z_1\overline{z}_1-z_2\overline{z}_2)^4
\\ \nonumber
&&+90 (z_1\overline{z}_2)^2(\overline{z}_1z_2)^2
(z_1\overline{z}_1-z_2\overline{z}_2)^2
\\ \nonumber
&&-20 (z_1\overline{z}_2)^3(\overline{z}_1z_2)^3.
\end{eqnarray}
In this  invariant polynomial we introduce  the 
Klein correspondence  eq. \ref{e36} and rewrite it in terms of 
$(\xi, \eta, \zeta)$.
We obtain the invariant homogeneous polynomial of degree 6
\begin{eqnarray}
\label{e40}
&&{\cal K}'(\xi,\eta,\zeta)=
\\ \nonumber
&&-42(1/2)^5((\xi+i\eta)^5\zeta+\zeta(\xi-i\eta)^5)
\\ \nonumber
&&+\zeta^6
\\ \nonumber
&&-30(1/2)^2 (\xi+i\eta)(\xi-i\eta)\zeta^4
\\ \nonumber
&&+90(1/2)^4 (\xi+i\eta)^2(\xi-i\eta)^2\zeta^2
\\ \nonumber
&&-20(1/2)^6 (\xi+i\eta)^3(\xi-i\eta)^3.
\end{eqnarray}
{\bf Lemma 6: The polynomial eq. \ref{e40} is an ${\cal H}_3$-invariant  
of degree 6 in $(\xi,\eta, \zeta)$}: 
The polynomial ${\cal K}'$ eq. \ref{e39}  
by construction is invariant under the binary icosahedral group 
${\cal H}_3$ acting on the complex coordinates. Moreover, the 
real coordinates  $(\xi, \eta, \zeta)$  eq. \ref{e36} carry a standard 
three-dimensional
orthogonal irrep of the icosahedral group (5-fold axis along 3-axis). 
Therefore the 
polynomial eq. \ref{e40} is also an invariant of degree  $6$ in 
$(\xi, \eta, \zeta)$
with respect to this irrep.

\section{Subduction of irreps for the pair $SO(4,R)>{\cal H}_3$.}

The group of deck transformations in the geometry of section 2
acts on $S^3 \sim SU(2,C)$ 
as the binary icosahedral group ${\cal H}_3$ from the right. 
We wish to find  on $S^3$  the homogeneous solutions of the Laplace
equation \ref{e27a} which belong to the identity irrep $D^{\alpha_0}$
of ${\cal H}_3$. This amounts to decomposing given irreps of a group $G$
into  irreps of a subgroup $G>H$, a process called subduction of irreps and 
reciprocal to induction. Actually we shall subduce the bases of irreps 
rather than the irreps themselves. For the  pair $SO(4,R)>{\cal H}_3$, by Lemma 2 we
can refine this analysis to $SO(4,R)>SU^r(2,C)>{\cal H}_3$. 
The fixed irreps
$D^j \times D^j$ of $SO(4,R)$ subduce the single irrep $D^j$ of $SU^r(2,C)$,
and so it remains to subduce the irreps in $SU^r(2,C)>{\cal H}_3$. 
This subduction has been studied by Cesare and Del Duca \cite{CE}. 
They distinguish between  fermionic and bosonic irreps of ${\cal H}_3$, 
corresponding to odd and even values of $\lambda=2j$. 
They give recursive expressions for the decomposition under ${\cal H}_3$ of all
irreps of $SU(2,C)$ and provide corresponding projection operators.

In the following sections we shall develop from Klein's 
fundamental invariant eq. \ref{e38} an
alternative and powerful  operator tool for
subducing  and quantizing an orthonormal and 
complete basis of eigenstates on $M$.

For the full subduction $SO(4,R)>{\cal H}_3$ we recall that 
${\cal H}_3$ commutes with $SU^l(2,C)$ so that we are free to choose 
the only free representation  label $m'$  of the latter group. For  
$m'=j$, which corresponds to the simple analytic 
spherical harmonics eq. \ref{e34},  or for any $m': -j\leq m'\leq j$, the subduction of the basis of spherical harmonics eq. \ref{e34}
reads
\begin{equation}
\label{e41} 
D^j_{j,\alpha}= \sum_m D^j_{j,m}(z,\overline{z})c_{m,\alpha},\;
-j\leq m'\leq j,
\end{equation}
\begin{equation}
\label{e42} 
D^j_{m',\alpha}= \sum_m D^j_{m',m}(z,\overline{z})c_{m,\alpha},\;
-j\leq m'\leq j.
\end{equation}
The coefficients $c_{m,\alpha}$ are yet to be determined but must be independent
of $m'$, since  we could pass from eq. \ref{e41} to eq. \ref{e42} by application 
of the left lowering operator $L^l_-$.
This procedure, exemplified by 
{\bf Table 1} for $\alpha=\alpha_0$,  reduces the search for spherical harmonics invariant under
${\cal H}_3$ by a factor $(2j+1)$.

{\bf Lemma 7: Classification of ${\cal H}_3$-invariant  polynomials 
by eigenvalues of $L^l_3$}:  For given $\lambda=2j$, the spherical harmonics which belong 
to a fixed irrep of ${\cal H}_3$ can be grouped into sets eq. \ref{e42} whose $(2j+1)$
members are orthogonal and 
distinguished by the eigenvalues $m',\; -j\leq m'\leq j$ of $L^l_3$.

\section{Irreducible  ${\cal H}_3$ states and 
the generalized Casimir operator ${\cal K}$ for $SU(2,C)>{\cal H}_3$.}

Casimir operators like $\Lambda^2$ are used for fixing 
the irreps of a given Lie group.
To distinguish and label the subduction 
of irreps in $G>H$, 
we follow the paradigm given by  Bargmann and Moshinsky \cite{BA} and
construct in the enveloping algebra of $SO(4,R)$ a generalized Casimir operator
${\cal C}$ associated to the group/subgroup pair $G>H$.
This Lie algebraic operator technique  was applied in \cite{BA} to 
$SU(3,C)>SO(3,R)$ and 
in \cite{KR4} pp. 263-8 to the continuous/discrete  pair 
$O(3,R)>D^{[3,1]}(S_4)$.  

A generalized Casimir operator must have the following properties:

(i) ${\cal C}$ must commute with the Casimir operators of the group $G$ and

(ii) ${\cal C}$  must be hermitian and invariant under the subgroup $H$ but not under 
the full group $G$.

Once we have found this operator for the groups $SU^r(2,C)>{\cal H}_3$,
by standard symmetry arguments its modes transforming with any fixed 
irrep of $SU(2,C)$ must fall into subsets of degenerate 
eigenstates  transforming 
with fixed irreps $D^{\alpha}$ of ${\cal H}_3$. 

Condition (i) is fulfilled 
by any operator-valued polynomial ${\cal P}$ in the generators of $SU^r(2,C)$.
The set of these polynomials forms the enveloping algebra of $SU^r(2,C)$.
The generators when taken as $(L^r_1, L^r_2, L^r_3)$  
transform linearly according to eq. \ref{e35}  under $SU^r(2,C)$ 
as the vector $(\xi,\eta, \zeta)$.

Condition (ii) requires the polynomial to be invariant under ${\cal H}_3$.
We would like  to get invariance by substituting
$(L^r_1,L^r_2,L^r_3)$   in  the invariant 
polynomial ${\cal K}'$ of degree $6$ 
from eq. \ref{e40}. But, as these  generators  
don't commute,
a naive substitution of $L$-components  into the polynomial eq. \ref{e40}
of degree $6$ 
does not guarantee  the invariant transformation property under ${\cal H}_3$.

The transformation property for any operator-valued polynomial 
${\cal P}(A_1,A_2,\ldots)$ 
is maintained if, after naive substitution, we symmetrize it by adding  
all polynomials obtained from any permutation of the operators 
$(A_1,A_2,\ldots)$ involved, 
and dividing by the 
number of permutations. The abelianization of this symmetrized
operator-valued polynomial clearly would reconstruct the polynomial  
in commuting vector components and its transformation property.
We denote the operation of  symmetrization by the
symbol $Sym({\cal P})$. 

{\bf Lemma 8: The ${\cal H}_3$-invariant operator ${\cal K}$}: 
The generalized Casimir operator ${\cal K}$ 
for the group/subgroup 
chain $SU^r(2,C) > {\cal H}_3$ is the hermitian polynomial operator
\begin{eqnarray}
\label{e43}
&&{\cal K}(L^r_1, L^r_2, L^r_3):=
\\ \nonumber
&&Sym({\cal K}'_{(\xi,\eta, \zeta) \rightarrow (L^r_1, L^r_2, L^r_3)})=
\\ \nonumber
&&-42 (1/2)^5 Sym(L_+^5L_3+L_3L_-^5)
\\ \nonumber
&&+Sym(L_3^6)
\\ \nonumber
&&-30 (1/2)^2 Sym(L_+L_-L_3^4)
\\ \nonumber
&&+90 (1/2)^4 Sym(L_+^2L_-^2L_3^2)
\\ \nonumber
&&-20 (1/2)^6 Sym(L_+^3L_-^3).
\end{eqnarray}
On the right-hand side of eq. \ref{e43} and in what follows 
we drop  for simplicity the 
upper index for the right action.
The hermitian property is either manifest or obtained upon symmetrization.
We observe that, in the standard $|jm>$ basis, 
the only off-diagonal terms of ${\cal K}$ are the first two.
The other four terms of ${\cal K}$ can be expressed as  
polynomials in $L^2, L_3$ and so
are diagonal. Algebraic expressions for  the operations $Sym$ of symmetrizations 
in eq. \ref{e43} are given in the Appendix.

\section{The spectrum of ${\cal K}$ and the quantization of $M$
by its eigenstates.}

To see if the operator ${\cal K}$ completely resolves the  subduction
of irreps in $SO(4,R)>{\cal H}_3$ we must explore its eigenvalues and
eigenstates. 
Since ${\cal K}$ is ${\cal H}_3$-invariant, its eigenstates on $\tilde{M}=S^3$ 
carry irreps $D^{\alpha}$ of
${\cal H}_3$. The operator ${\cal K}$ classifies and by its eigenvalues 
quantizes  the irrep $D^{\alpha}$ of ${\cal H}_3$ 
for fixed $j$ on $S^3$. 
The states in general live on the universal 
cover $\tilde{M}=S^3$ and must have an additional degeneracy corresponding to
the dimension $|\alpha|$ of the irrep $D^{\alpha}$. The combined eigenspaces 
of ${\cal K}$ 
span the same linear space as the spherical 
harmonics on $S^3$, but now   transforming 
under irreps  
$D^{\alpha}$ of ${\cal H}_3$.

To diagonalize ${\cal K}$ we apply Lemma 5 in eigenspaces of fixed $\lambda=2j$
and for $m'=j$. The corresponding subspace ${\cal L}^j_j$ 
has dimension $2j+1$ and an analytic  basis $|jm>$ eq. \ref{e34}
w.r.t. the right action  of $SU^r(2,C)$. The only off-diagonal elements 
arise from
the first two terms of ${\cal K}$ eq. \ref{e43}, 
while the other ones are diagonal in this scheme.

We now take full advantage of Klein's expressions eq. \ref{e23}  
for the elements of ${\cal H}_3$. 
Since the off-diagonal  part of ${\cal K}$ links the basis states modulo $5$, 
we split the values of $m,\;  -j\leq m \leq j$ as 
$m \equiv \mu\; {\rm modulo}\; 5$ 
and the subspace ${\cal L}^j_j$  into orthogonal subspaces 
${\cal L}^j_{j,\mu}$ of fixed $\mu$. 
Within a subspace ${\cal L}^j_{j,\mu}$, the matrix of ${\cal K}$ is 
tridiagonal, moreover with non-zero off-diagonal entries.
>From these properties it is easily  shown \cite{BA} that, 
within ${\cal L}^j_{j,\mu}$, 
the spectrum of ${\cal K}$
is non-degenerate. This implies that the $|\alpha|$ degenerate eigenstates
of ${\cal K}$ belonging to the same irrep $D^{\alpha}$ are completely 
distinguished by the index $\mu$. The  other degeneracy
of ${\cal K}$, resulting from its commuting with $SU^l(2,C)$, 
is resolved by diagonalization of $L^l_3$ with eigenvalue $m'$,
as exemplified  in {\bf Table 1}.

Among the irrep $D^{\alpha}$ modes on $\tilde{M}=S^3$ is the subset of proper 
eigenstates for the topological 
Poincar\'{e} 3-manifold $M$. They belong exclusively to the identity irrep of 
${\cal H}_3$, which we denote by  $D^{\alpha_0} \equiv  1$. Since any  value taken by 
such a polynomial function on $S^3$ is repeated on all 
copies of $M$ under ${\cal H}_3$,
the domain of these states can be  uniquely restricted from $\tilde{M}$
to the Poincar\'{e} 3-manifold $M$.
For these invariant eigenstates we get sharper selection rules
of the subspaces ${\cal L}^j_{j,\mu}$.
We have taken the elements of ${\cal H}_3$ in the setting due to Klein \cite{KL}
eq. \ref{e23}.
The binary preimage of the icosahedral 5-fold rotation around the 3-axis 
is generated by the element $S$ from this equation. 
For the eigenstates belonging to $D^{\alpha_0}$, invariance in particular
under the element $S \in {\cal H}_3$ implies
that they can occur only in the subspaces 
$({\cal L}^j_{m',\mu},\mu=0)$.
These subspaces appear only for $2j=$ even, and so there can be 
no ${\cal H}_3$-invariant eigenmodes of $M$ with $2j=$ odd.
Due to the non-degeneracy of ${\cal K}$ on these subspaces, 
any two ${\cal H}_3$-invariant 
eigenstates on the same subspace $({\cal L}^j_{m',\mu},\mu=0)$ must differ
in their eigenvalues.

{\bf Theorem 1: The ${\cal H}_3$-invariant operator ${\cal K}$ eq. \ref{e43}
quantizes the 
Poincar\'{e}'s dodecahedral 3-manifold $M$}: A complete set of 
invariant eigenmodes on the dodecahedral 
Poincar\'{e} topological manifold $M$ is given by those
eigenstates with eigenvalue $\kappa$ of ${\cal K}$ 
which belong to the identity irrep $D^{\alpha_0}$
of ${\cal H}_3$. These eigenstates    
occur only in the orthogonal subspaces 
$({\cal L}^j_{m',\mu},\mu=0,\, 2j={\rm even},\; -j\leq m'\leq j)$. For fixed $j$, they are
of degree $2j$ and are quantized  by the eigenvalue 
${\bf \kappa}$ of ${\cal K}$. In the subspace 
$({\cal L}^j_{(m',\mu)}, m'=j,\,\mu=0)$, 
the eigenstates are non-degenerate and 
are homogeneous polynomials 
analytic in $(z_1,z_2)$ as in eq. \ref{e34} 
and similar to Klein's $f_k$ eq. \ref{e38}. 
The $2j$ partners with the same eigenvalue $\kappa$
and eigenvalue $m'=j-1,\ldots, -j$ of $L^l_3$ are obtained 
by multiple application of the lowering operator $L^l_-$,
as given in  {\bf Table 1}.

Theorem 1 verifies the preview given in the introduction.
The different topologies of $\tilde{M}=S^3$ versus $M$ give rise to different
eigenmodes: On $\tilde{M}$, the eigenmodes are all the spherical
harmonics of eq. \ref{e32}. To pass to the eigenmodes of $M$,
one must select from them the ${\cal H}_3$-invariant ones, 
which have a unique restriction to $M$. The number of invariant 
eigenmodes for fixed $j$ can also be found from \cite{CE}, the 
eigenmodes become eigenstates of ${\cal K}$.

{\bf Lemma 9: No eigenmodes of {M} exist for $\lambda=2j <12$}.\\
{\em Proof}: The irreps $D^j,\; j=(1,2)$ of $SU^r(2)$ remain 
irreducible when subduced to ${\cal H}_3$,
\cite{CE} p. 5, and so cannot yield invariant eigenmodes. 
For $j=(3,4,5)$, the diagonalization of ${\cal K}$ 
in section 10 shows only degenerate
and no invariant eigenstates at all.

\section{Diagonalization of ${\cal K}$\\
for the subspaces  ${\cal L}^j,\; j=1,\ldots,6$.}

First of all we observe that in the subspaces 
${\cal L}^j,\;j=1,2$, the operator ${\cal K}$
gives vanishing results. The reason is as follows: 
The polynomials ${\cal K}'$ eq. \ref{e39} as well as $f_k$ 
under $SU^r(2,C)$ transform according to the irrep $D^j,\; j=6$.
It then follows from the construction eq. \ref{e41} 
that the operator ${\cal K}$
under $SU^r(2,C)$ transforms as part of a tensor operator 
${\cal K}={\cal K}^q$ of rank $q=6$. 
>From standard selection rules for tensor operators, 
non-vanishing of its matrix  elements 
$<j_2m_2|{\cal K}^6|j_1m_1>$ requires $j_1+j_2\geq 6\geq |j_1-j_2|$
which excludes $j_1=j_2=(1,2)$. 
The irreps of ${\cal H}_3$
for $j=(1,2)$ can  be analyzed independent of ${\cal K}$
but provide no invariant modes, see Lemma 9.  

As an illustration of  ${\cal K}$ and its diagonalization we consider
the subspaces ${\cal L}^{j},\; j=3,4,5,6$.
Within each subspace ${\cal L}^j_{j,\mu}$, the submatrix ${\cal K}^{\mu}$ of ${\cal K}$, 
the diagonalizing matrix
$V^{\mu}$, 
and the diagonal form ${\cal K}^{\mu, diag}$ fulfill
\begin{equation}
\label{e43a}
{\cal K}^{\mu}\cdot V^{\mu}=V^{\mu}\cdot {\cal K}^{\mu, diag}.
\end{equation}
As a survey we give in Table 2 the eigenvalues $\kappa$ and 
multiplicities $|\alpha|$  
in the form $\kappa^{|\alpha|}$ as a function of $j$.

\begin{table}
$
\begin{array}{ll}
j        & \kappa^{|\alpha|}\\
3    &(-225)^3,(\frac{675}{4})^4\\
4    &(\frac{7875}{4})^4,(-1575)^5\\
5    &(-\frac{23625}{2})^3,(7875)^3,(\frac{4725}{2})^5\\
6    &(-51975)^1, (-\frac{51975}{2})^3, (23625)^4,(\frac{14175}{2})^5
\end{array}
$
\caption{Eigenvalues $\kappa$ of ${\cal K}$ and their multiplicities 
$|\alpha|$ in the form $\kappa^{|\alpha|}$ for $j=3,4,5,6$.}
\end{table}
We give  the  submatrices of ${\cal K}$
according to eq. \ref{e43a} in closed algebraic form in Tables 3-6.
The extension of the diagonalization to $j>6$ offers no problem.

\newpage
\begin{table}

\begin{eqnarray}
\label{e44a}
&& j=3,\; \kappa^{|\alpha|}= (-225)^3,(\frac{675}{4})^4.
\\ \nonumber
&& \mu=0: m=0
\\ \nonumber
&&{\cal K}^{0}=\left[-225\right],\;
V^{0}=\left[1\right],\;
{\cal K}^{0,diag}=\left[-225\right]
\\ \nonumber
&& \mu=1: m=1
\\ \nonumber
&&{\cal K}^{1}=\left[\frac{675}{4}\right],\;
V^{1}=\left[1\right],\;
{\cal K}^{1,diag}=\left[\frac{675}{4}\right]
\\ \nonumber
&&\mu=2:\; m=(-3,2):
\\ \nonumber
&&{\cal K}^{2}=\left[
\begin{array}{rr}
\frac{45}{4}&315\sqrt{\frac{3}{8}}\\
315\sqrt{\frac{3}{8}} &-\frac{135}{2}
\end{array}
\right],\;
\\ \nonumber
&&V^{2}=\left[
\begin{array}{rr}
-\sqrt{\frac{2}{5}}&\sqrt{\frac{3}{5}}\\
\sqrt{\frac{3}{5}} &\sqrt{\frac{2}{5}}
\end{array}
\right],\;
{\cal K}^{2,diag}=\left[
\begin{array}{rr}
-225&\\
&\frac{675}{4}
\end{array}
\right],
\\ \nonumber
&&\mu=3:\; m=(-2,3):
\\ \nonumber
&&{\cal K}^{3}=\left[
\begin{array}{rr}
-\frac{135}{2}&-315\sqrt{\frac{3}{8}}\\
-315\sqrt{\frac{3}{8}}&\frac{45}{4}
\end{array}
\right],\;
\\ \nonumber
&&V^{3}=\left[
\begin{array}{rr}
\sqrt{\frac{3}{5}}&-\sqrt{\frac{2}{5}}\\
\sqrt{\frac{2}{5}} &\sqrt{\frac{3}{5}}
\end{array}
\right],\;
{\cal K}^{3,diag}=\left[
\begin{array}{rr}
-225&\\
&\frac{675}{4}
\end{array}
\right],
\\ \nonumber
&&\mu=4: m=-1
\\ \nonumber
&&{\cal K}^{4}=\left[\frac{675}{4}\right],\;
V^{4}=\left[1\right],\;
{\cal K}^{4,diag}=\left[\frac{675}{4}\right]
\end{eqnarray}
\caption{Submatrices of ${\cal K}$ following eq. \ref{e43a} for $j=3$.}
\end{table}

\begin{table}
\begin{eqnarray}
\label{e44b}
&& j=4,\; \kappa^{|\alpha|}= (\frac{7875}{4})^4,(-1575)^5.
\\ \nonumber
&&\mu=0: m=0
\\  \nonumber
&&{\cal K}^{0}=\left[-1575\right],\;
V^{0}=\left[1\right],\;
{\cal K}^{0,diag}=\left[-1575\right]
\\ \nonumber
&&\mu=1:\; m=(-4,1):
\\ \nonumber
&&{\cal K}^{1}=\left[
\begin{array}{rr}
315&945\sqrt{\frac{7}{2}}\\
945\sqrt{\frac{7}{2}}&\frac{315}{4}
\end{array}
\right],\;
\\ \nonumber
&&V^{1}=\left[
\begin{array}{rr}
2\sqrt{\frac{2}{15}}&-\sqrt{\frac{7}{15}}\\
\sqrt{\frac{7}{15}} &2\sqrt{\frac{2}{15}}
\end{array}
\right],\;
{\cal K}^{1,diag}=\left[
\begin{array}{rr}
\frac{7875}{4}&\\
&-1575
\end{array}
\right],
\\ \nonumber
&&\mu=2:\; m=(-3,2):
\\ \nonumber
&&{\cal K}^{2}=\left[
\begin{array}{rr}
-\frac{5355}{4}&945\sqrt{\frac{7}{8}}\\
945\sqrt{\frac{7}{8}} &\frac{3465}{2}
\end{array}
\right],\;
\\ \nonumber
&&V^{2}=\left[
\begin{array}{rr}
\sqrt{\frac{1}{15}}&-\sqrt{\frac{14}{15}}\\
\sqrt{\frac{14}{15}} &\sqrt{\frac{1}{15}}
\end{array}
\right],\;
{\cal K}^{2,diag}=\left[
\begin{array}{rr}
\frac{7875}{4}&\\
&-1575
\end{array}
\right],
\\ \nonumber
&&\mu=3:\; m=(-2,3):
\\ \nonumber
&&{\cal K}^{3}=\left[
\begin{array}{rr}
\frac{3465}{2}&-945\sqrt{\frac{7}{8}}\\
-945\sqrt{\frac{7}{8}}&-\frac{5355}{4}
\end{array}
\right],\;
\\ \nonumber
&&V^{3}=\left[
\begin{array}{rr}
-\sqrt{\frac{14}{15}}&\sqrt{\frac{1}{15}}\\
\sqrt{\frac{1}{15}} &\sqrt{\frac{14}{15}}
\end{array}
\right],\;
{\cal K}^{3,diag}=\left[
\begin{array}{rr}
\frac{7875}{4}&\\
&-1575
\end{array}
\right],
\\ \nonumber
&&\mu=4:\; m=(-1,4):
\\ \nonumber
&&{\cal K}^{4}=\left[
\begin{array}{rr}
\frac{315}{4}&-945\sqrt{\frac{7}{2}}\\
-945\sqrt{\frac{7}{2}} &315
\end{array}
\right],\;
\\ \nonumber
&&V^{4}=\left[
\begin{array}{rr}
-\sqrt{\frac{7}{15}}&\sqrt{\frac{8}{15}}\\
\sqrt{\frac{8}{15}} &\sqrt{\frac{7}{15}}
\end{array}
\right],\;
{\cal K}^{4,diag}=\left[
\begin{array}{rr}
\frac{7875}{4}&\\
&-1575
\end{array}
\right].
\end{eqnarray}
\caption{Submatrices of ${\cal K}$ following eq. \ref{e43a} for $j=4$.}
\end{table}

\begin{table}
\begin{eqnarray}
\label{e44c}
&& j=5,\; \kappa^{|\alpha|}= (-\frac{23625}{2})^3,(7875)^3,(\frac{4725}{2})^5.
\\ \nonumber
&&\mu=0:\; m=(-5,0,5):
\\ \nonumber
&&{\cal K}^{0}=\left[
\begin{array}{rrr}
\frac{4725}{2}&\frac{4725\sqrt{7}}{2}&0\\
\frac{4725\sqrt{7}}{2}&-6300&-\frac{4725\sqrt{7}}{2}\\
0&-\frac{4725\sqrt{7}}{2}&\frac{4725}{2}
\end{array}
\right],
\\ \nonumber
&&V^{0}=\left[
\begin{array}{rrr}
-\sqrt{\frac{7}{50}}&-\frac{3}{5}       &\sqrt{\frac{1}{2}}\\
\sqrt{\frac{36}{50}} &-\frac{\sqrt{7}}{5}&0\\
\sqrt{\frac{7}{50}}&\frac{3}{5}        &\sqrt{\frac{1}{2}}
\end{array}
\right],\;
{\cal K}^{0,diag}=\left[
\begin{array}{rrr}
-\frac{23625}{2}&&\\
&7875&\\
&&\frac{4725}{2}
\end{array}
\right],
\\ \nonumber
&&\mu=1:\; m=(-4,1):
\\ \nonumber
&&{\cal K}^{1}=\left[
\begin{array}{rr}
-7560&\frac{2835\sqrt{21}}{2}\\
\frac{2835\sqrt{21}}{2}&-1890
\end{array}
\right],\;
\\ \nonumber
&&V^{1}=\left[
\begin{array}{rr}
-\sqrt{\frac{7}{10}}&\sqrt{\frac{3}{10}}\\
\sqrt{\frac{3}{10}} &\sqrt{\frac{7}{10}}
\end{array}
\right],\;
{\cal K}^{1,diag}=\left[
\begin{array}{rr}
-\frac{23625}{2}&\\
&\frac{4725}{2}
\end{array}
\right],
\\ \nonumber
&&\mu=2:\; m=(-3,2):
\\ \nonumber
&&{\cal K}^{2}=\left[
\begin{array}{rr}
\frac{9135}{2}&2205\sqrt{\frac{3}{2}}\\
2205\sqrt{\frac{3}{2}}      &5670
\end{array}
\right],\;
\\ \nonumber
&&V^{2}=\left[
\begin{array}{rr}
\sqrt{\frac{2}{5}}&-\sqrt{\frac{3}{5}}\\
\sqrt{\frac{3}{5}} &\sqrt{\frac{2}{5}}
\end{array}
\right],\;
{\cal K}^{2,diag}=\left[
\begin{array}{rr}
7875&\\
&\frac{4725}{2}
\end{array}
\right],
\\ \nonumber
&&\mu=3:\; m=(-2,3):
\\ \nonumber
&&{\cal K}^{3}=\left[
\begin{array}{rr}
5670&-2205\sqrt{\frac{3}{2}}\\
-2205\sqrt{\frac{3}{2}}&\frac{9135}{2}
\end{array}
\right],\;
\\ \nonumber
&&V^{3}=\left[
\begin{array}{rr}
-\sqrt{\frac{3}{5}}&\sqrt{\frac{2}{5}}\\
\sqrt{\frac{2}{5}} &\sqrt{\frac{3}{5}}
\end{array}
\right],\;
{\cal K}^{3,diag}=\left[
\begin{array}{rr}
7875&\\
&\frac{4725}{2}
\end{array}
\right],
\\ \nonumber
&&\mu=4:\; m=(-1,4):
\\ \nonumber
&&{\cal K}^{4}=\left[
\begin{array}{rr}
-1890&-\frac{2835\sqrt{21}}{2}\\
-\frac{2835\sqrt{21}}{2} &-7560
\end{array}
\right],\;
\\ \nonumber
&&V^{4}=\left[
\begin{array}{rr}
\sqrt{\frac{3}{10}}&-\sqrt{\frac{7}{10}}\\
\sqrt{\frac{7}{10}} &\sqrt{\frac{3}{10}}
\end{array}
\right],\;
{\cal K}^{4,diag}=\left[
\begin{array}{rr}
-\frac{23625}{2}&\\
&\frac{4725}{2}
\end{array}
\right],
\end{eqnarray}
\caption{Submatrices of ${\cal K}$ following eq. \ref{e43a} for $j=5$.}
\end{table}

\begin{table}
\begin{eqnarray}
\label{e44d}
&& j=6,\kappa^{|\alpha|}= (-51975)^1, (-\frac{51975}{2})^3, (23625)^4,
(\frac{14175}{2})^5.
\\ \nonumber
&&\mu=0,\; m=(-5,0,5):\\
\nonumber
&&{\cal K}^{0}=\left[
\begin{array}{rrr}
-\frac{51975}{2}&\frac{4725\sqrt{77}}{2}&0\\
\frac{4725\sqrt{77}}{2}&-18900&-\frac{4725\sqrt{77}}{2}\\
0&-\frac{4725\sqrt{77}}{2}& -\frac{51975}{2}
\end{array}
\right],
\\ \nonumber
&&V^{0}=\left[
\begin{array}{rrr}
-\sqrt{\frac{7}{25}}&\sqrt{\frac{1}{2}}&-\sqrt{\frac{11}{50}}\\
\sqrt{\frac{11}{25}}&0&-2\sqrt{\frac{7}{50}}\\
\sqrt{\frac{7}{25}}&\sqrt{\frac{1}{2}}&\sqrt{\frac{11}{50}}  
\end{array}
\right],\;
{\cal K}^{0, diag}=\left[
\begin{array}{rrr}
-51975&&\\
&-\frac{51975}{2}&\\
&&\frac{14175}{2}
\end{array}
\right].
\\ \nonumber
&&\mu=1,\; m=(-4,1,6):\\
\nonumber
&&{\cal K}^{1}=\left[
\begin{array}{rrr}
3780&\frac{19845\sqrt{3}}{2}&0\\
\frac{19845\sqrt{3}}{2}&-9450&-6615\sqrt{\frac{11}{2}}\\
0&-6615\sqrt{\frac{11}{2}}& 10395
\end{array}
\right],
\\ \nonumber
&&V^{1}=\left[
\begin{array}{rrr}
-\sqrt{\frac{11}{50}}&-\sqrt{\frac{6}{25}}&3\sqrt{\frac{3}{50}}\\
\sqrt{\frac{33}{50}}&-2\sqrt{\frac{2}{25}}&\sqrt{\frac{1}{50}}\\
\sqrt{\frac{6}{50}}&\sqrt{\frac{11}{25}}&\sqrt{\frac{22}{50}}  
\end{array}
\right],\;
{\cal K}^{1, diag}=\left[
\begin{array}{rrr}
-\frac{51975}{2}&&\\
&23625&\\
&&\frac{14175}{2}
\end{array}
\right].
\\ \nonumber
&&\mu=2,\; m=(-3,2):
\\ \nonumber
&&{\cal K}^{2}=\left[
\begin{array}{rr}
\frac{40635}{2}&6615\\
6615&10395
\end{array}
\right],\;
\\ \nonumber
&&V^{2}=\left[
\begin{array}{rr}
2\sqrt{\frac{1}{5}}&-\sqrt{\frac{1}{5}}\\
\sqrt{\frac{1}{5}}&2\sqrt{\frac{1}{5}}
\end{array}
\right],\;
{\cal K}^{2, diag}=\left[
\begin{array}{rr}
23625&\\
&\frac{14175}{2}
\end{array}
\right]
\\ \nonumber
&&\mu=3,\; m=(-2,3):
\\ \nonumber
&&{\cal K}^{3}=\left[
\begin{array}{rr}
10395&-6615\\
-6615&\frac{40635}{2}
\end{array}
\right],\;
\\ \nonumber
&&V^{3}=\left[
\begin{array}{rr}
-\sqrt{\frac{1}{5}}&2\sqrt{\frac{1}{5}}\\
2\sqrt{\frac{1}{5}}&\sqrt{\frac{1}{5}}
\end{array}
\right],\;
{\cal K}^{3, diag}=\left[
\begin{array}{rr}
23625&\\
&\frac{14175}{2}
\end{array}
\right].
\\ \nonumber
&&\mu=4,\; m=(-6,-1,4):
\\ \nonumber
&&{\cal K}^{4}=\left[
\begin{array}{rrr}
10395&6615\sqrt{\frac{11}{2}}&0\\
6615\sqrt{\frac{11}{2}}&-9450&-\frac{19845\sqrt{3}}{2}\\
0&-\frac{19845\sqrt{3}}{2}&3780 
\end{array}
\right],
\\ \nonumber
&&V^{4}=\left[
\begin{array}{rrr}
-\sqrt{\frac{6}{50}}&-\sqrt{\frac{11}{25}}&\sqrt{\frac{22}{50}}\\
\sqrt{\frac{33}{50}}&-2\sqrt{\frac{2}{25}}&-\sqrt{\frac{1}{50}}\\
\sqrt{\frac{11}{50}}&\sqrt{\frac{6}{25}}&\sqrt{\frac{27}{50}}  
\end{array}
\right],\;
{\cal K}^{4, diag}=\left[
\begin{array}{rrr}
-\frac{51975}{2}&&\\
&23625&\\
&&\frac{14175}{2}
\end{array}
\right]
\end{eqnarray}
\caption{Submatrices of ${\cal K}$ following eq. \ref{e43a} for $j=6$.}
\end{table}

\newpage
The structures in  Tables 3-6 are algebraic and display the typical properties of the spectrum and eigenstates
of ${\cal K}$ as described in Theorem 1: 
Within each subspace ${\cal L}^j_{\mu}$ there are no
degenerate eigenvalues. Eigenvalues are repeated only in subspaces
$\mu' \neq \mu$. The only single eigenvalue ${\bf \kappa}= -51975$ occurs in
${\cal L}^6_{6,0}$ and 
determines a single  eigenstate on the Poincar\'{e} manifold. 
The corresponding eigenstate when written in terms of the normalized
polynomials eq. \ref{e34} as
\begin{eqnarray}
\label{e45}
&&\left[
\frac{z_1z_2^{11}}{\sqrt{11!}}\cdot (-\sqrt{\frac{7}{25}})
+\frac{z_1^6z_2^6}{\sqrt{6!6!}}\cdot \sqrt{\frac{11}{25}}
+\frac{z_1^{11}z_2}{\sqrt{11!}}\cdot\sqrt{\frac{7}{25}}
\right]
\\ \nonumber
&&\equiv -f_k(z_1,z_2)\sqrt{\frac{7}{25\cdot 11!}},
\end{eqnarray}
proves to be  proportional to 
Klein's analytic invariant eq. \ref{e38}. So the invariant operator ${\cal K}$ 
eq. \ref{e43}, derived from Klein's invariant, reproduces this 
invariant as an eigenstate quantized by its eigenvalue. 
This crucial result confirms the consistency 
of the present approach.
Moreover the states given in Table 1
up to normalization are (part of) the 12 $m'$-partners of Klein's invariant
and belong to the same eigenvalue of ${\cal K}$. These form the lowest 
degree eigenmodes of $M$.

\section{Discussion.}

We compare the analysis with recent work on cosmic topology.
Multiply connected topologies for  cosmology have become a 
field of intense study
\cite{LA}, \cite{LE}.  
The authors of   \cite{LU}, \cite{WE}  
propose in particular the Poincar\'{e} 3-manifold $M$ as a candidate 
for the space part  of the cosmos. 
In their terminology it belongs to the single-action
manifolds, corresponding to the right action of the  
group of deck transformations. In Lemma 2 we prove this right action from the
gluing prescription of \cite{SE1},\cite{SE2} for the dodecahedral 3-manifold.

With the goal to expand the temperature fluctuations 
of the cosmic microwave background (CMB),
the eigenmodes of $M$
and of similar 3-manifolds 
are studied  
in \cite{LEH} by a ghost, an averaging, and by a projection method.

There are some conceptual differences: 
The authors of \cite{LEH} on p. 4687 speak of
eigenstates of the Laplacian, whereas Weeks \cite{WE} p. 615 characterizes 
the modes as homogeneous harmonic polynomials of degree $k$ 
solving the Laplace equation, $\Delta P=0$.
The latter notion agrees with the spherical harmonics
according to section 5,  which by eqs. \ref{e27},\ref{e27a}, \ref{e28} 
diagonalize the 
Casimir operator of $SO(4,R)$, with   
$\lambda=2j$ playing the role of $k$. 
The  ghost method of \cite{LEH}
looks for a  restriction 
of eigenmodes of the universal covering
$S^3$ to those of $M$,
which agrees with the reasoning  
given in section 1, but  no general expressions for the eigenmodes are given.
Weeks, \cite{WE} p. 615, points out  the need for an accurate 
and efficient computation of the eigenmodes. 
These properties are  provided by the present operator and quantization 
method.
 
Lachi\`{e}ze-Rey \cite{LA2} discusses the eigenmodes of $M$ in terms
of modified spherical harmonics on $S^3$. His basis explicitly reduces 
the  cyclic group $Z_5$ as in Klein's analysis.  
The results on eigenmodes in 
\cite{LA2} Table 1 are given in numerical and not in algebraic form. 

The selection rules of  eigenmodes of $M$ versus those of  $\tilde{M}=S^3$
are emphasized  by Weeks \cite{WE}. However, dodecahedral   
quadrupole and octupole modes $l=2,3$, 
as discussed in \cite{LU},
\cite{WE}, are  in conflict with the selection rule 
$k=2j \geq 12$  on $M$.
All the selection rules for eigenmodes can be read off  already from the 
irrep  subduction rules for $SU(2,C)>{\cal H}_3$ given in \cite{CE}, 
complemented by  the multiplicity $(2j+1)$ arising from ${\cal H}_3$
commuting with $SU^l(2,C)$.
Ikeda \cite{IK} gives the lowest degree of a non-vanishing 
eigenmode of $M$ as $k=12$ with multiplicity $13$. These values agree
with $j=6$ and multiplity $2j+1=13$ of the present  analysis.

\section{Conclusion.}

The eigenmodes of the Poincar\'{e} dodecahedral topological 
3-manifold $M$ are characterized by Lie algebraic operator techniques 
as eigenstates with eigenvalues.  
Guided by homotopy, the group of deck transformations  and related 
Coxeter groups, by 
F. Klein's fundamental invariant polynomial, and by representation theory, 
a hermitian generalized 
Casimir operator ${\cal K}$ for the
group/subgroup  subduction $SO(4,R)>SU^r(2,C)>{\cal H}_3$
is constructed. Its eigenstates are obtained from homogeneous  
polynomials, analytic in two complex variables $(z_1,z_2)$.
The degeneracies in the spectrum of ${\cal K}$
are completely resolved. The proper selection rules for
passing from eigenstates on the universal covering $\tilde{M}=S^3$
to eigenstates on $M$ arise from the spectrum of ${\cal K}$.
The basis of eigenstates of ${\cal K}$ is well
suited for the expansion of observables like the temperature
fluctuation of the CMB.

The present Lie algebraic operator techniques from representation theory are not 
restricted
to $M$, they can be developed for
other models and groups \cite{LEH} considered in cosmic topology.  
For example one could think of
the orbifold associated with the Coxeter group eq. \ref{e12} consisting of
the fundamental simplex 
for this group described in section 4.

Hyperbolic counterparts of the Poincar\'{e} dodecahedral 3-manifold
are the Weber-Seifert dodecahedral 3-manifold \cite{SE1} and variants
of it given by Best \cite{BE}. Similar 
methods from group theory, including a hyperbolic Coxeter group,
apply to the Weber-Seifert 3-manifold, compare \cite{KR3}.

\section*{Acknowledgment.}
It is a pleasure to thank  T. Kramer 
for substantial help in the algebraic computations. The invitation by  M. Moshinsky, G. S. Pogosyan,
L. E. Vicent and K. B. Wolf
to present this work at the
25th ICGTMP, Cocoyoc, Mexico 2004 is gratefully acknowledged. 

\section*{Appendix: Explicit symmetrization of the operator ${\cal K}$.}

The operation $Sym$ of symmetrization of an operator-valued polynomial is 
well defined, it can be used to  compute the matrix, eigenvectors and eigenvalues
of ${\cal K}$ eq. \ref{e43}. Here we develop an 
alternative efficient method for obtaining its matrix 
in the $|jm>$ scheme, based on the representation of the Lie algebra of
$SU^r(2,C)$ and its commutators eq. \ref{e31}. 

We begin with the first two terms in the $H_3$-invariant operator 
${\cal K}$ eq. \ref{e43} and obtain
with the help of the commutators
\begin{eqnarray}
\label{A1}
&&Sym(L_+^5L_3)=\frac{1}{6}\sum_{\nu=0}^5L_+^{5-\nu}L_3L_+^{\nu} 
\\ \nonumber 
&&=\frac{1}{6}\left[6L_+^5L_3+15L_+^5\right],
\\ \nonumber
&&Sym(L_-^5L_3)=\frac{1}{6}\sum_{\nu=0}^5L_-^{\nu}L_3L_-^{5-\nu} 
\\ \nonumber 
&&=\frac{1}{6}\left[6L_3L_-^5+15L_-^5\right],
\\ \nonumber 
&&Sym(L_+^5L_3)+Sym(L_-^5L_3)
\\ \nonumber
&&=\frac{1}{6}\left[6L_+^5L_3+15L_+^5+6L_3L_-^5+15L_-^5\right].
\end{eqnarray}
The next term of ${\cal K}$ offers no problem, we find  
$Sym(L_3^6)=L_3^6$. 

The following three terms of ${\cal K}$ have equal powers in 
$(L_+,L_-)$. Therefore they can be expressed as polynomial
functions of the commuting operators $(L^2,L_3)$. 
Consider the term of ${\cal K}$ quadratic in $(L_+,L_-)$. These two operators
can appear in two orders which for short we denote as 
\begin{eqnarray}
\label{A1a}
&&(AB)=
\\ \nonumber
&&(+-),(-+).
\end{eqnarray} 
Using the commutator and the Casimir  
invariant $L^2$, we get the well-known results
\begin{equation}
\label{A2}
L_+L_-=L^2-L_3(L_3-1),\; 
L_-L_+=L^2-L_3(L_3+1).
\end{equation}
Both operators are diagonal with respect to states labelled by
$(j,m)$. It proves convenient to pass to the matrix elements
by writing
\begin{eqnarray}
\label{A3}
&&a(m):=\langle jm|L_+L_-|jm\rangle=j(j+1)-m(m-1),
\\ \nonumber
&&a(m+1)=\langle jm|L_-L_+|jm\rangle=j(j+1)-m(m+1),
\\ \nonumber
&& a(m) = 0\; {\rm if}\;  m<-j,\;  m>j.
\end{eqnarray}
The coefficients $a(m)$ eq. \ref{A3}
will appear in the following equations of this section.
For short we suppress their dependence on the  fixed irrep label $j$.  
In the full term of ${\cal K}$ quadratic in $(L_+,L_-)$  
we must now insert four powers of 
$L_3:= C$. In $Sym$ there appear $15$ monomial terms of the order
$\ldots A\ldots B\ldots$. We order them as
\begin{eqnarray}
\label{A4}
&&(C^4AB), (C^3ACB), (C^3ABC), (C^2AC^2B), (C^2ACBC),
\\ \nonumber
&&(C^2ABC^2), (CAC^3B), (CAC^2BC), (CACBC^2), (CABC^3),
\\ \nonumber
&&(AC^4B), (AC^3BC), (AC^2BC^2), (ACBC^3), (ABC^4).
\end{eqnarray}
When we pass to the diagonal matrix elements of the terms in eq. \ref{A4}, 
the powers of $L_3=C$ contribute quartic expressions in $m$ whose 
values depend on the choice of $(AB)$ and on the order eq. \ref{A4} 
of insertion.
A straightforward evaluation in terms of $a(m)$ eq. \ref{A3}, $m$
yields for the full sum of all these $30$ terms
\begin{eqnarray}
\label{A5}
&&\langle jm|Sym(L_+L_-L_3^4)|jm\rangle
\\ \nonumber
&&=\frac{1}{30}a(m) (15m^4-20m^3+15m^2-6m+1)
\\ \nonumber
&&+\frac{1}{30}a(m+1)(15m^4+20m^3+15m^2+6m+1).
\end{eqnarray}  
The evaluation of the monomials in ${\cal K}$ of power $4$ in 
$(L_+,L_-)$ 
proceeds in the same fashion. First we order the powers
of $(L_+, L_-)$ in short-hand notation as
\begin{eqnarray}
\label{A6}
&&(ABDE)= 
\\ \nonumber
&&(++--), (+-+-), (+--+),
\\ \nonumber 
&&(-++-), (-+-+), (--++).
\end{eqnarray}
For any fixed order $(ABDE)$ chosen from eq. \ref{A6}, the two additional 
powers of $C=L_3$ can be inserted
according to the $15$ terms
\begin{eqnarray}
\label{A7}
&&(C^2ABDE), (CACBDE), (CABCDE), (CABDCE),(CABDEC),
\\ \nonumber
&&(AC^2BDE), (ACBCDE), (ACBDCE),(ACBDEC),(ABC^2DE),
\\ \nonumber
&&(ABCDCE), (ABCDEC),(ABDC^2E), (ABDCEC), (ABDEC^2).
\end{eqnarray}
The powers of $L_3$ contribute quadratic expressions  in $m$,
and the full sum of  $90$ monomial terms,
after summing in each one  over the $15$ terms eq. \ref{A7},
reduces in terms  of $a(m)$ from eq. \ref{A3}  to 
\begin{eqnarray}
\label{A8}
&&\langle jm|Sym(L_+^2L_-^2L_3^2)|jm\rangle
\\ \nonumber
&&=\frac{1}{90}a(m-1)a(m)(15m^2-24m+11)
\\ \nonumber
&&+\frac{1}{90}a(m)a(m)(15m^2-12m+3)
\\ \nonumber
&&+\frac{1}{90}a(m)a(m+1)(15m^2+1)
\\ \nonumber
&&+\frac{1}{90}a(m+1)a(m)(15m^2+1)
\\ \nonumber
&&+\frac{1}{90}a(m+1)a(m+1)(15m^2+12m+3)
\\ \nonumber
&&+\frac{1}{90}a(m+2)a(m+1)(15m^2+24m+11).
\end{eqnarray}
We keep all $6$ terms in correspondence to eq. \ref{A6}.

Finally we evaluate from ${\cal K}$ the sum of monomials of power $6$ 
in $(L_+,L_-)$. The $20$ orderings  can be abridged as
\begin{eqnarray}
\label{A9}
&&(ABDEFG)=
\\ \nonumber
&& (+++---), (++-+--), (++--+-),(++---+),
\\ \nonumber
&& (+-++--),(+-+-+-), (+-+--+),(+--++-),
\\ \nonumber
&&  (+--+-+), (+---++),(-+++--), (-++-+-),
\\ \nonumber
&& (-++--+), (-+-++-), (-+-+-+), (-+--++),
\\ \nonumber
&& (--+++-), (--++-+), (--+-++), (---+++).
\end{eqnarray}
The evaluation yields in the order of eq. \ref{A9}
\begin{eqnarray}
\label{A10}
&&\langle jm|Sym(L_+^3L_-^3)|jm\rangle = \frac{1}{20}
\\ \nonumber
&&\begin{array}{llllll}
(a(m-2) &\cdot a(m-1)&\cdot a(m)  &+a(m-1)&\cdot a(m-1)&\cdot a(m)\\
+a(m-1) &\cdot a(m)  &\cdot a(m)  &+a(m-1)&\cdot a(m)  &\cdot a(m+1)\\
+a(m)   &\cdot a(m-1)&\cdot a(m)  &+a(m)  &\cdot a(m)  &\cdot a(m)\\
+a(m)   &\cdot a(m)  &\cdot a(m+1)&+a(m)  &\cdot a(m+1)&\cdot a(m)\\
+a(m)   &\cdot a(m+1)&\cdot a(m+1)&+a(m)  &\cdot a(m+2)&\cdot a(m+1)\\
+a(m+1) &\cdot a(m-1)&\cdot a(m)  &+a(m+1)&\cdot a(m)  &\cdot a(m)\\
+a(m+1) &\cdot a(m)  &\cdot a(m+1)&+a(m+1)&\cdot a(m+1)&\cdot a(m)\\
+a(m+1) &\cdot a(m+1)&\cdot a(m+1)&+a(m+1)&\cdot a(m+2)&\cdot a(m+1)\\
+a(m+2) &\cdot a(m+1)&\cdot a(m)  &+a(m+2)&\cdot a(m+1)&\cdot a(m+1)\\
+a(m+2) &\cdot a(m+2)&\cdot a(m+1)&+a(m+3)&\cdot a(m+2)&\cdot a(m+1)).\\
\end{array}
\end{eqnarray}
One can easily convert 
all expectation values back into operators by inserting
the commuting operators $(L^2, L_3)$ in eqs. \ref{A5},\ref{A8},\ref{A10}.
It can be  recognized that the terms in ${\cal K}$ of highest power
in $(L_+,L_3,L_-)$ correspond to the commutative invariant eq. \ref{e40}.
All other terms reflect the non-commutative structure of ${\cal K}$.

\end{document}